\documentclass[12pt]{article}

\usepackage{jheppub}
\usepackage{amsmath}
\usepackage{amssymb}
\usepackage{epsfig,amssymb,amsmath,psfrag,hyperref}
\usepackage[dvipsnames]{xcolor}
\usepackage{float} 
\usepackage{bbm} 
\usepackage{mathtools} 
\usepackage[enableskew]{youngtab} 
\usepackage{ytableau} 
\usepackage{slashed} 
\numberwithin{equation}{section}
\usepackage{caption}
\usepackage{booktabs}
\usepackage{mwe}
\usepackage{gensymb}
\usepackage{ulem}
\usepackage{cleveref}
\usepackage[toc,page]{appendix}
\usepackage{xcolor}

\usepackage{graphicx}

\usepackage{multirow, graphicx,amssymb,url,mathrsfs,amsmath}
\usepackage{eucal,wrapfig,boxedminipage,setspace,subfigure}
\usepackage{amsxtra,amstext,latexsym,dsfont,xcolor}

\def\IR{{\hbox{{\rm I}\kern-.2em\hbox{\rm R}}}}
\def\IB{{\hbox{{\rm I}\kern-.2em\hbox{\rm B}}}}
\def\IN{{\hbox{{\rm I}\kern-.2em\hbox{\rm N}}}}
\def\IC{\,\,{\hbox{{\rm I}\kern-.59em\hbox{\bf C}}}}
\def\IZ{{\hbox{{\rm Z}\kern-.4em\hbox{\rm Z}}}}
\def\IP{{\hbox{{\rm I}\kern-.2em\hbox{\rm P}}}}
\def\IH{{\hbox{{\rm I}\kern-.4em\hbox{\rm H}}}}
\def\ID{{\hbox{{\rm I}\kern-.2em\hbox{\rm D}}}}

\newcommand{\beq}{\begin{equation}}
\newcommand{\eeq}{\end{equation}}
\newcommand{\bea}{\begin{eqnarray}}
\newcommand{\eea}{\end{eqnarray}}

\usepackage{cleveref}

\newcommand{\myslash} [1] {#1 \kern-.5em/}

\begin{document}

$\left. \right.$   \vspace{-18cm}

\title{$\left. \right.$ \vspace{17cm} \\ Designer bubble walls in a holographic Weyl semi-metal with magnetic field  }

\author[a]{Nick Evans,}
\author[a]{Wanxiang Fan}


\affiliation[a]{School of Physics \& Astronomy and STAG Research Centre, University of Southampton, Highfield,  Southampton  SO$17$ $1$BJ, UK.}

\emailAdd{n.j.evans@soton.ac.uk; w.fan@soton.ac.uk}

\abstract{We study bubble walls in the holographic D3/probe D7 system that is dual to strongly coupled quark dynamics. We work to construct holographic descriptions of bubble wall junctions where we can tune the bubble wall height and pressure difference in a theory at zero temperature and density. We study the system in the presence of a background axial vector field $b$ (associated with the $z$ direction)  that induces a Weyl semi-metal, massless phase and a perpendicular background magnetic field $B_x$ which favours mass generation.  We find a first order transition line in the mass-$B_x$  plane, at fixed $b$, ending at a critical point. We present some preliminary solutions of PDEs that describe the motion of one dimensional bubble walls in this theory, with a stationary initial condition - large pressure differences accelerate the wall to the speed of light whilst when the pressure difference is small the wall slumps to an interpolating solution. We also take the first steps to include temperature and see evidence of thermal drag slowing the wall motion. Slump configurations at finite temperature show some signs of a back pressure wave against the wall motion.}

\maketitle

\setcounter{page}{1}\setcounter{footnote}{0}

\newpage

\section{Introduction}

A first order phase transition in the early Universe would generate the formation of bubbles \cite{Hindmarsh:2020hop,DeCurtis:2022hlx}. As the bubbles grow and collide gravitational wave signals may be generated \cite{Chala:2018ari,Caldwell:2022qsj}. There has been considerable interest in this phenomena in the light of current and future gravitational wave detectors. A key component is to compute the speed at which bubble walls move \cite{Moore:1995ua,Ai:2021kak,Laurent:2022jrs, Wang:2022txy, Janik:2022wsx}. If the system undergoing the phase transition were to be strongly coupled (in the spirit of the QCD phase transition albeit with a first order transition) then calculation becomes yet harder \cite{Ares:2021nap}. 

A number of groups have used holography \cite{Maldacena:1997re,Witten:1998qj} to study bubble motion in such strongly coupled systems \cite{Ares:2021ntv,Janik:2021jbq,Bigazzi:2021ucw,Bea:2022mfb,Bea:2021zsu}. The standard assumption is that in steady state motion of the bubble wall, the friction term counters the pressure difference across the wall. Thus, for example, as one heats up a system towards a first order transition one expects the friction to grow and the pressure difference across the wall to drop and at some point they will begin to compete. The work in \cite{Bea:2022mfb,Bea:2021zsu} provides simulations of the resultant motion and \cite{Bigazzi:2021ucw, Henriksson:2021zei} computes the friction analytically in a probe brane setting. On the other hand \cite{Janik:2022wsx} has seen cases in simulations where the pressure equalises around the bubble wall and a pressure wave instead travels separately and apart form the wall. This suggests that it would be useful to have further holographic examples with more control parameters and simulations of such systems. 

In this work our goal is to find a holographic description of a theory with bubble walls where we can dial the parameters of the bubble wall and independently control the relative strength of the thermal bath providing the friction. This requires finding a first order transition in parameter space that ends at a critical point. At the critical point the barrier height between vacua will fall to zero for example. In this paper we work to construct just such a set of theories using the D3/probe D7 holographic dual \cite{Karch:2002sh,Kruczenski:2003be,Erdmenger:2007cm} of quark physics and do first simulations of those bubble walls' motions.  We include temperature (and hope in the future to consider density) and watch the effect of the background bath impeding the wall's motion. 

The holographic model we work with is the D3/probe D7 system that is a well known \cite{Karch:2002sh,Kruczenski:2003be,Erdmenger:2007cm} and simple description of quenched ${\cal N}=2$ supermultiplets of quarks in a background of ${\cal N}=4$ supersymmetric glue. The benefit of the probe system will be that we can study time dependent solutions for the bubble wall movement without changing the thermal background geometry. This greatly reduces the difficulty of solving the partial differential equations (PDEs), at least in some cases, so that solutions can be extracted simply using the \verb  |NDSolve| function from \verb|MATHEMATICA|. On the other hand \verb|MATHEMATICA| is an imperfect tool that can overwhelm desktops for longer time runs and at some parameter values, so we present these results as preliminary, leaving a full scan of the parameter space, in a more bespoke computational setting, for future work.

Let's first address our goal to  construct a first order transition in parameter space that ends at a critical point. To realise this we introduce three parameters to the quark physics. They are: a source from a U(1)$_A$ background gauge field $b$ (associated with the $z$ diretion) that makes the system a Weyl semi-metal \cite{BitaghsirFadafan:2020lkh}; a quark mass $M_{UV}$; and a perpendicular background magnetic field $B_x$ \cite{Filev:2007gb}. We will summarize our findings for this theory (which are of interest purely in the study of the Weyl semi-metal phase) before returning to bubble wall motion later.

Weyl and Dirac semi-metals are materials in which an insulator, when in the presence of some background field, develops light degrees of freedom that conduct strongly \cite{Yan:2016euz,Armitage:2017cjs,ZahidHasan:2017fhz,Gao:2018xwm,Liu:2018djq}. The simplest, weakly coupled, relativistic model is a massive fermion, $\psi$, (which at low energies can not be excited) with a coupling to an axial gauge field \cite{Colladay:1998fq,Grushin:2012mt}

\begin{equation}
\mathcal{L}=\Bar{\psi}(i \gamma^{\mu} \partial_{\mu}-M_{UV}+A^5_j \gamma^j \gamma^5)\psi 
\end{equation}

where the spacetime index $\mu=t,x,y,z$, $\gamma^{\mu}$ are the Dirac matrices, and $\gamma^{5}$ is defined as $i\gamma^{t}\gamma^{x}\gamma^{y}\gamma^{z}$, $j=1,2,3$ is the spatial index, but can be chosen as in a single spatial direction e.g. $z$ due to the rotational symmetry\cite{BitaghsirFadafan:2020lkh}, $M_{UV}$ is the bare quark mass. 
When the axial gauge field develops a vev, for example in the $z$ direction, $A_z^5 = b/2$, the energy eigenvalues of the system are (here one applies the adjoint of the Dirac equation to the Dirac equation itself and diagonalises the resultant Klein Gordon operator)

\footnotesize
\begin{equation}
    E^2 = k_x^2 + k_y^2 + \left(\frac{b}{2} \pm \sqrt{k_z^2+M_{UV}^2} \right)^2
\end{equation}
\normalsize
There are two Weyl spinor, zero energy states with \footnotesize $(k_x, k_y, k_z)= (0,0,\pm \sqrt{(b/2)^2 - M_{UV}^2})$ \normalsize provided $b/2 > M_{UV}$, which can now be excited at low energies.

There has been considerable interest in seeking holographic models of such behaviour \cite{Landsteiner:2015lsa,Landsteiner:2015pdh,
Copetti:2016ewq,Liu:2018spp,Landsteiner:2019kxb} and the D3/probe D7 system has been one system in which it has been realised \cite{BitaghsirFadafan:2020lkh}. In \cite{BitaghsirFadafan:2020lkh} it was shown that switching on a vev for the axial U(1) gauge field of the fermion fields indeed reproduces a phase transition at low quark mass (relative to $b$) to a phase with a massless IR.  The D7 brane mebeddings are shown in Figure 1, as we will discuss, and the large $b$ phase are those that touch the origin of the AdS space, consistent with the presence of an ungapped mode. The phase transition is first order and occurs at $M_{UV}/bR^2 = M_{UV}^*/bR^2 = 0.46$ where $R$ is the AdS radius.

That first order transition is the starting point for our investigation. 
In this paper we will in addition perturb that semi-metal system by switching on a magnetic field perpendicular to the axial gauge field (the effect of a magnetic field in an alternative holographic model is in \cite{Bruni:2023ife}). This is directly interesting in that system because there are two competing effects - the parameter $b$ favours a theory with gapless degrees of freedom. On the other hand a magnetic field $B_x$ is known to spontaneously generate a mass gap in the holographic system \cite{Filev:2007gb}. At some fixed value of $b$ one might expect a transition in the mass $M_{UV}/bR^2$ versus magnetic field $B_x$ plane as one or other parameter dominates. In fact though, as we will see, the magnetic field acts to generate a mass gap in the semi-metal phase the instant it is non-zero. The transition when $M_{UV} \simeq M_{UV}^*$ remains as a first order transition at low $B_x$ but as $B_x$ increases the energy barrier between the two vacua at the transition shrinks before the transition terminates at a critical point. We show explicitly that there is a massless scalar field at the critical point.  Note the case of a magnetic field parallel to the axial field ansatz is more complicated since there are additional terms effecting the embedding from the Wess Zumino action term - we leave this case for future exploration.

More generally though one can view adding $B_x$ as simply adding a parameter that breaks the conformal invariance of the strongly coupled system. The effect is to allow the strong interactions to generate a mass gap in the field theory. This is the behaviour one would expect from QCD or a Nambu-Jona-Lasinio  model \cite{Nambu:1961fr} of strong interactions.  Thus we can view the size of $B_x$ as somewhat analogous to the scale of strong coupling in the system. To see the semi-metal transition we need the parameter $b$ to be large or of order that strong coupling scale. Thus generically we expect a strongly coupled system that is gapped to resist the semi-metal phase.

 It is important to note at this point that in \cite{Kharzeev:2011rw} the possibility of a striped phase instability was investigated in the D3/probe D7 system with magnetic field. The Wess-Zumino term in the presence of a magnetic field introduces a coupling between the phase of the embedding and the gauge field on the D7 world volume. They argued that this term would lead to an instability. In this paper we neglect this instability and assume these fields are zero. The solutions we find are self-consistent solutions of the equations of motion but may not be the global minima. Our main justification for this is that we want a model of domain walls between different homogeneous vacua since this is the usual cosmological scenario considered. Including the WZ term and looking at these striped instabilities would be an interesting extension of our work but here we neglect it.

We also include temperature into this system - the semi-metal massless phase at $B_x=0$ immediately moves to a thermal plasma phase in the presence of even infinitesimal temperature for all $M_{UV}<M_{UV}^*$ as shown in \cite{BitaghsirFadafan:2020lkh}.  As $B_x$ rises there is an IR gap of order $B_x$ below $M_{UV}^*$ and there is then a first order transition when $T$  becomes of order that gap where the stable mesons melt into a thermal plasma of quasi-normal mode excitations \cite{Hoyos-Badajoz:2006dzi}. As $T$ grows the thermal transition grows along the line of the first order transitions at $B_x=0$ until the meson melting transition has swallowed the semi-metal transition line. At yet higher $B_x$ or $T$ the transition becomes the familiar meson melting transition of the magnetic field theory at $b=0$ \cite{Albash:2007bk}.

We have now constructed the ground works for our studies of bubble walls. The D3/probe D7 system with $b,M_{UV},B_x$ has within it's parameter space regions with multiple local minima of the potential. We can dial the pressure (action) difference between two local minima by moving towards or away from the first order transition line where the two vacua are degenerate. We can change the height of the potential  hill between the two local minima by choosing parameters near or far from the critical point (where that difference vanishes). We therefore can set up bubble walls on the boundary between pieces of space in each local minima with controlled properties. We can then  add a thermal bath via a temperature in the background geometry and observe its effects on the wall motion.

In this paper we just look at the motion of one dimensional domain walls. We make a simple ansatz that interpolates between two local minima of the potential. We are able to track this motion using the simple \verb  |NDSolve| function from \verb|MATHEMATICA| for a wide set of initial conditions (we monitor energy conservation of the configurations to ensure the solutions are sensible). On the other hand  \verb|MATHEMATICA| is not optimised to these problems and can overwhelm the processor on a desktop machine for some cases. 
Generically an initial condition describing a wall moves with time in our solutions due to the pressure difference across the wall. At zero temperature the walls rapidly accelerate to moving at the speed of light. The only exception is when we start with a configuration with a small energy difference (in units of $M_{UV},b,B_x$, the parameters of our theory)  between the two vacuum and the meta-stable vacuum. Here the solutions prefer to slump into an interpolating configuration between the vacua and it is hard to describe the system as a moving domain wall. We report on some particular exemplars of this behaviour.

At finite temperature we report on two configurations, representative of a larger class of similar examples we have explored, that show novel behaviour beyond the $T=0$ set ups. Firstly, raising the temperature as much as we can we find that bubble wall motion, at least in the early stages, is slower than at $T=0$ providing evidence for drag force due to the plasma. The speed of motion is reduced to $0.7\,c$ (at least in the early period of motion) in the case we highlight. Secondly we report on configurations with smaller pressure differences across the wall which evolve to states with two separated pressure walls moving in opposite directions. This suggests an intermediate state to those observed in \cite{Janik:2022wsx} where the pressure fully equilibriates at the wall but a back pressure wave travels through the plasma.

These preliminary results support the programme presented here that allows us to introduce temperature (and we hope in the future chemical potential) into simulations of the dynamics of strongly coupled bubble walls using holography. We leave the   wider  systematic study of initial configurations and understanding longer time evolution for future work (likely using numerical techniques beyond  \verb|MATHEMATICA| ).

We develop the model we study in Sections 2-5 including identifying the massless quark anti-quark bound state  expected at the critical point. In sections 6-7 we present the results of simulations of the bubble wall motion at zero temperature and in the presence of temperature respectively.

\section{The Weyl Semi-Metal Transition in the D3/D7 System}

The dynamics of $N_f$ coincident D7-brane probes embedded in the $AdS_5\times S^5$ geometry is encoded in the Dirac-Born-Infeld
(DBI) and Wess-Zumino (WZ) action \cite{Karch:2002sh,Kruczenski:2003be,Erdmenger:2007cm}
\begin{align}\label{Equation: DBI action and WZ action}
    S_{D7}=-N_{f}T_{D7} \int d^{8}\xi \sqrt{-\text{det}(P[G]+F)}+ \frac{1}{2}N_f T_{D7} \int P[C_4]\wedge F\wedge F,
\end{align}
where  $T_{D7}$ is the D7-brane tension, $\xi^{a}$ with $a=1,2,...,8$ are the world volume coordinates, $P[G]$ and $P[C_4]$ denotes the pullback of the bulk metric and four-form induced on the D7-brane, and the $F=dA$ is the field strength of corresponding gauge field $A$. 

The D3-branes' and D7-branes' world coordinates are parameterised as 
\footnotesize
\begin{align}
\begin{tabular}{ c|c c c c c c c c c c }
       & t & x & y & z & $\rho$ &$\beta_1$&$\beta_2$&$\beta_3$& L & $\phi$ \\ 
 \hline
    D3 & x & x & x & x &   &         &         &         &   &        \\  
    D7 & x & x & x & x & x & x       & x       & x       &   &   
\end{tabular}
\end{align}
\normalsize
where the x implies the world volume coordinates, $\rho$ and the $3-$sphere angles $\beta_i$ parametrize the polar coordinates on the D7 brane in the bulk. $L,\phi$ are the polar coordinates in the $2-$plane transverse to the D7.

We will work generically in the 
the $AdS_{5}-Schwarzchild$ metric dual to an ${\cal N}=4$ gauge theory (at finite temperature when $\rho_h \neq 0$). We write the metric in coordinates appropriate to embed the D7 brane \cite{Babington:2003vm}
\begin{align} \label{metricT2}
    ds^2=\frac{r^2}{R^2} \left( -\frac{g^2(r)}{h(r) }dt^2+h(r)(dx^2+dy^2+dz^2)\right)+\frac{R^2}{r^2} \left( d \rho ^2+\rho ^2 ds^{2}_{ S^3}+dL^2+L^2 d \phi^{2} \right),
\end{align}
where $R$ is the AdS radius.
The four form is given by
\begin{align} \label{c4}
    C_4=\frac{r^4}{R^4} h^2(r) dt\wedge dx\wedge dy \wedge dz -\frac{R^4 \rho^4}{r^4} d\phi \wedge \omega(S^3),
\end{align}
where 
\begin{align}
    r^2=\rho^2+L^2, \hspace{0.5cm} g(r)=1-\frac{r^4_{H}}{r^4},\hspace{0.5cm}  h(r)=1+\frac{r^4_{H}}{r^4}
\end{align}
where $r_H$ is the horizon radius, the corresponding Hawking temperature, dual to the field theory's temperature, is
\begin{align}
    T = \frac{\sqrt{2}}{\pi} \frac{r_H}{R^2}.
\end{align}

The usual ansatz for the D7 embedding, dictated by the symmetries of the problem, is $L=L(\rho)$. In addition here we will set $\phi(z)=b z$ which corresponds to the background 
axial U(1) gauge field that triggers the semi-metal phase \cite{BitaghsirFadafan:2020lkh}. The DBI action is given by
\begin{align}\label{eqns:DBI action without B}
    S_{D7}=-\mathcal{N} \text{vol}\left( \mathbb{R}^{1,3}\right)\int d\rho \rho^3 g(r) h(r) \sqrt{\left( 1+\frac{R^4 b^2 L^2}{h(r)(\rho^2+L^2)^2}\right)(1+L'^2)}
\end{align}
where $\mathcal{N}=2\pi^2 N_{f} T_{D7}$. At zero temperature, the action is given by setting $r_H=0$ and hence $g=h=1$.  Note here the WZ term has played no role in the action \cite{BitaghsirFadafan:2020lkh} - we discuss this in more detail below when we also have magnetic field present.

\begin{center}
\includegraphics[width=13cm,height=8cm]{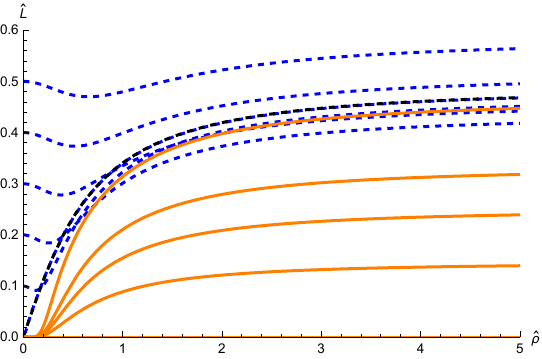} 

\textit{Figure 1: The solutions for $\hat{L}(\rho)$ as a function of $\hat{M}_{UV}$ in the model at $\hat{B}_x=\hat{T}=0$. The dashed solutions are regular in the IR ($\hat{L}'(0)=0$) and correspond to massive theories  with stable mesonic fluctuations. The solid lines are solutions that end on the D3 brane at $\hat{\rho}=\hat{L}=0$ corresponding to the semi-metal phase at large $b/M_{UV}$. The dashed lines to the massive phase at small $b/M_{UV}$.}  
\end{center}

The asymptotic expansion of $L$ in the UV is 
\begin{align}\label{asymptotic expansion of L}
    L=M_{UV}+ \frac{C}{\rho^2}+\cdots
\end{align}
where $M_{UV}$ is the bare quark mass and $C$ is related to the dimesnion 3 quark condensate operator. 

Writing $L$ and $\rho$ in units of $R^2b$ for numerical work we find at $T=0$
\begin{align}
    S_{D7}=-\mathcal{N} \text{vol}\left( \mathbb{R}^{1,3}\right)(R^2b)^4\int d\hat{\rho}\, \hat{\rho}^3  \sqrt{\left( 1+\frac{\hat{L}^2}{(\hat{\rho}^2+\hat{L}^2)^2}\right)(1+\hat{L}'^2)}
\end{align}
where $\hat{L}'=\hat{L}'(\hat{\rho})$. For more details, see \autoref{app: rescaled quantities}.

The Euler Lagrange equation for $\hat{L}(\hat{\rho})$ is straight forwards to derive. It can be numerically solved using the \verb  |NDSolve| function from \verb|MATHEMATICA|. The method is essentially shooting: for the semi-metal phase we use IR boundary conditions  $\hat{L}(0)=0$ and vary the derivative to obtain a particular UV mass value; for the Minkowski massive phase we set $\hat{L}'(0)=0$ and vary $\hat{L}(0)$ again to achieve a particular UV mass value. 

We display the set of solutions for $\hat{L}(\hat{\rho})$ as a function of $M_{UV}/R^2b$ in Figure 1. The dashed solutions are regular in the IR ($\hat{L}'(0)=0$) and correspond to massive theories (the UV mass can be read off from the UV asymptotic value of $L$ - formally $m_q=M_{UV}/( 2 \pi \alpha' R^2 b)$)  with stable mesonic fluctuations. The solid lines are solutions that end on the D3 brane at $\hat{\rho}=\hat{L}=0$. These have been interpreted as the semi-metal phase with a massless IR. The two sets of solutions overlap indicating the presence of a first order phase transition - analysis of the free energy in \cite{BitaghsirFadafan:2020lkh} showed the transition was at $M_{UV}/R^2b=0.46$.

\section{Magnetic Field}

Our main focus in this section is to look at the effect on the Weyl semi-metal phase of an applied magnetic field. We will introduce a background magnetic field in the $x$ direction through the $F_{yz}$ component of the U(1) vector gauge field in the DBI D7 action \cite{Filev:2007gb} (or equivalently in the $y$ direction through $F_{xz}$). {\bf Note we work throughout in radial gauge $A_\rho=0$ (see for example eq.(2.21) in \cite{Sakai:2005yt}).}
The DBI action with $F_{yz}=B_x$ is
\vspace{-0.8cm} 

\footnotesize
\begin{align} \label{Baction}
    S_{D7}=-\mathcal{N} \text{vol}\left( \mathbb{R}^{1,3}\right)\int d\rho \,\rho^3 g(r) h(r) \sqrt{\left( 1+\frac{R^4 B_{x}^{2}}{h(r)^2 (\rho^2+L^2)^2}+\frac{R^4 b^2 L^2}{h(r)(\rho^2+L^2)^2}\right)(1+L'^2)}
\end{align} \normalsize
where $\mathcal{N}=2\pi^2 N_{f} T_{D7}$. 

 It is worth discussing the role of the WZ term at this point. The $C_4$ form enters via the second term in (\ref{c4}) which since, $d \phi = b dz$ with the ansatz for $\phi$, soaks up the $z$ and $S^3$ indices. The terms, which contain $b$ and $L$  are therefore contracted with one of the quadratic factors 
$F_{tx} \wedge F_{y\rho}$, $F_{ty} \wedge F_{x\rho}$ or $F_{t \rho} \wedge F_{xy}$. For our choice $B_x$, since these terms are quadratic in these fields they can all be set to zero consistently in the equations of motion. Had one switched on $B_z=F_{xy}$ this would not be the case and additional terms would result in a charge density in the model (see (7.9) of \cite{Landsteiner:2016led}). We concentrate on $B_x$ in this paper since it is productive and simpler (see Appendix A for the term that develops if $B_z$ is non-zero). There is one additional complication pointed out in \cite{Kharzeev:2011rw}. If one looks at fluctuations of $\phi$ in for example the $x$ direction then there is a term that couples that fluctuation quadratically to $A_t$ in the presence of the magnetic field (here the $C_4$ sucks up indices on the 3-sphere, and $x$ since $d \phi = (\partial_x \phi) dx$; the $B_x$ field uses indices $y,z$ and there is a coupling to $\partial_\rho A^t$) . Although it remains the case that one can self consistently set $A^t, \partial_x \phi$ to zero, when one allows their fluctuation, the spectrum for these states is tachyonic \cite{Kharzeev:2011rw}. This suggests there is a  lower energy vacuum with these fields switched on which is likely a striped phase. We will not include this instability in this paper, although it is interesting, because we wish to holographically study bubble walls between homogeneous vacuums. Our vacuums are likely metastable though in the full model therefore. It would be interesting to study this in the future, but hence forth we neglect the WZ term.

   The corresponding rescaled action to \eqref{Baction} is derived in \autoref{app: rescaled quantities}:
\footnotesize
\begin{align}
        &S_{D7}=-\mathcal{N}  \text{vol}\left( \mathbb{R}^{1,3}\right)(R^2 b)^4\int d\hat{\rho} \,\hat{\rho}^3 g h \sqrt{\left( 1+\frac{ \hat{B}_{x}^{2}}{h^2 (\hat{\rho}^2+\hat{L}^2)^2}+\frac{\hat{L}^2}{h(\hat{\rho}^2+\hat{L}^2)^2}\right)(1+\hat{L}'^2)},
\end{align}
\normalsize
where $\hat{B}_x=B_x/R^2b^2$, $g=g(\hat{\rho})$ and $h=h(\hat{\rho})$ for short.

In the case of $\hat{T}$ and $\hat{B}_x$ presented, the general UV asymptotic expansion takes the form:
\footnotesize
\begin{equation}
    L[\hat{\rho}] = \sum_{n=0}^{\infty} \left( c[n] + d[n] \log{(\hat{\rho})} \right) \hat{\rho}^{-n},
\end{equation}
\normalsize
where we find the coefficients:
\footnotesize
\begin{equation}\label{eqn:L UV expansion}
    L[\hat{\rho}] = \hat{M}_{UV} \left(1 - \frac{\log{\hat{\rho}}}{2\hat{\rho}^2}\right) + \frac{\hat{C}}{\hat{\rho}^2} + \frac{c[4]}{\hat{\rho}^4} + \frac{d[4]\log{\hat{\rho}}}{\hat{\rho}^4} + O\left(\frac{\log{\hat{\rho}}}{\hat{\rho}^6}\right),
\end{equation}
\normalsize
where
\footnotesize
\begin{equation*}
    d[4] = d[2]/8 = -\hat{M}_{UV}/16,
    \quad
    c[4] = \frac{1}{64} \left( 8\hat{C} + \hat{M}_{UV} (3 + 16\hat{B}_x^2 + 32\hat{M}_{UV}^2) \right).
\end{equation*}
\normalsize
We include the counter terms from \cite{BitaghsirFadafan:2020lkh} plus the $\hat{B}_x$ field introduces a new counter term which is $\hat{B}_x$ dependent. In total we have (note here $b=1$, $R=1$ and its factors resolve dimensional differences)
\footnotesize
\begin{equation} \label{ct}
   \hat{{\cal L}}_{\rm counter} = \frac{1}{2} \log (\hat{\rho}_{uv}) \left( \hat{M}_{UV}^2 + \hat{B}_x^2 \right) - \frac{1}{4} \left( \hat{M}_{UV}^2 \right) (2 \log (\hat{M}_{UV}) + 1) + \frac{\hat{\rho}_{uv}^4}{4},
\end{equation}
\normalsize
where $\hat{\rho}_{uv}$ is the UV cut off of the holographic radius.



We can now begin to explore the effects of the magnetic field on the Weyl semi-metal phase. In Figure 2, we show plots of $\hat{L}(\hat{\rho})$ (again found numerically by the shooting method) and varying $\hat{B}_x$—for the moment, we set $\hat{T} = 0$. Even at small $\hat{B}_x$, there are no longer solutions that end at $\hat{L} = \hat{\rho} = 0$. All solutions become of the 



\newpage

\begin{center}
       \includegraphics[width=6.7cm,height=4.8cm]{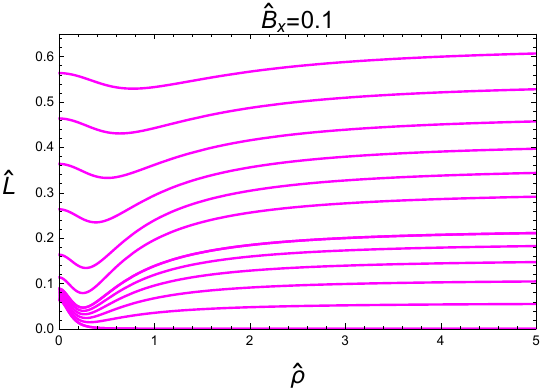}
       \includegraphics[width=6.7cm,height=4.8cm]{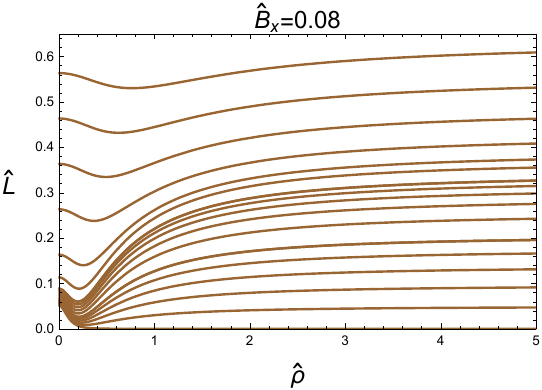}\\
        \includegraphics[width=6.7cm,height=4.8cm]{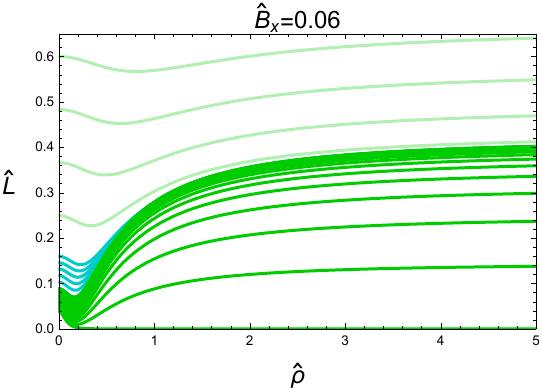}
        \includegraphics[width=6.7cm,height=4.8cm]{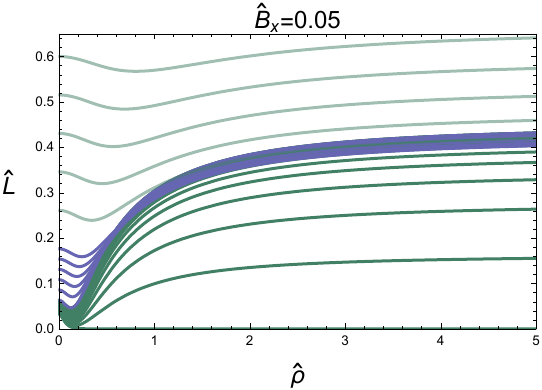}\\
        \includegraphics[width=6.7cm,height=4.8cm]{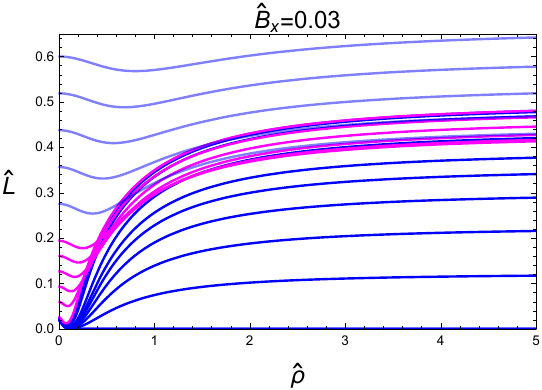}
        \includegraphics[width=6.7cm,height=4.8cm]{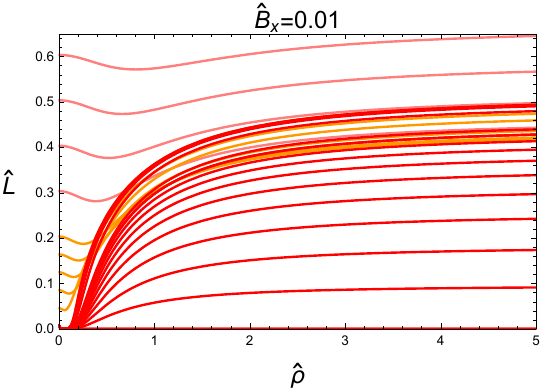}\\
        \includegraphics[width=8.2cm,height=5.2cm]{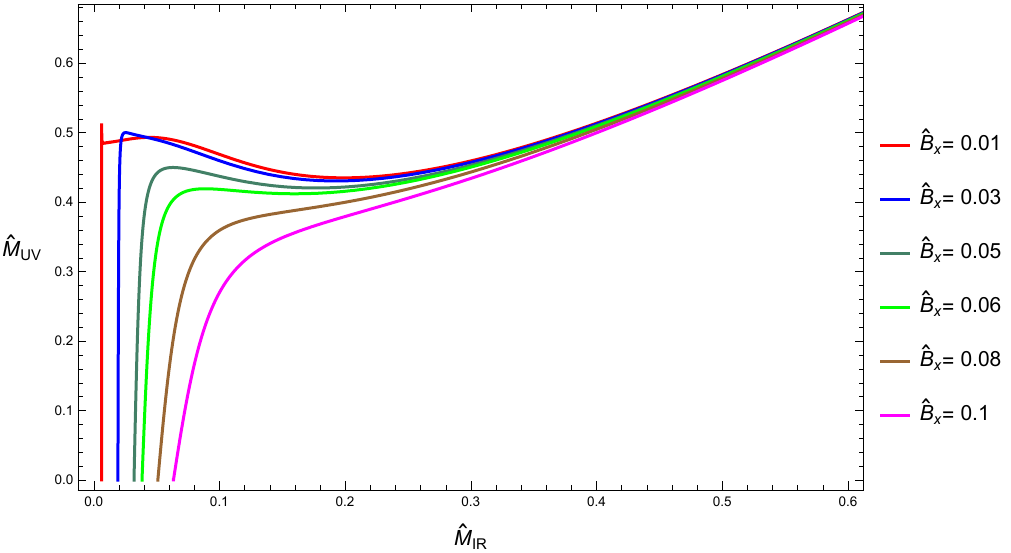}\\
        \textit{Figure 2: Embeddings $\hat{L}(\rho)$ for different $\hat{B}_x$ values. The lowest plot shows the UV mass vs IR mass for the embeddings as a function of $\hat{B}_x$. }   
        \end{center}

\begin{center}
        \includegraphics[width=12cm,height=7cm]{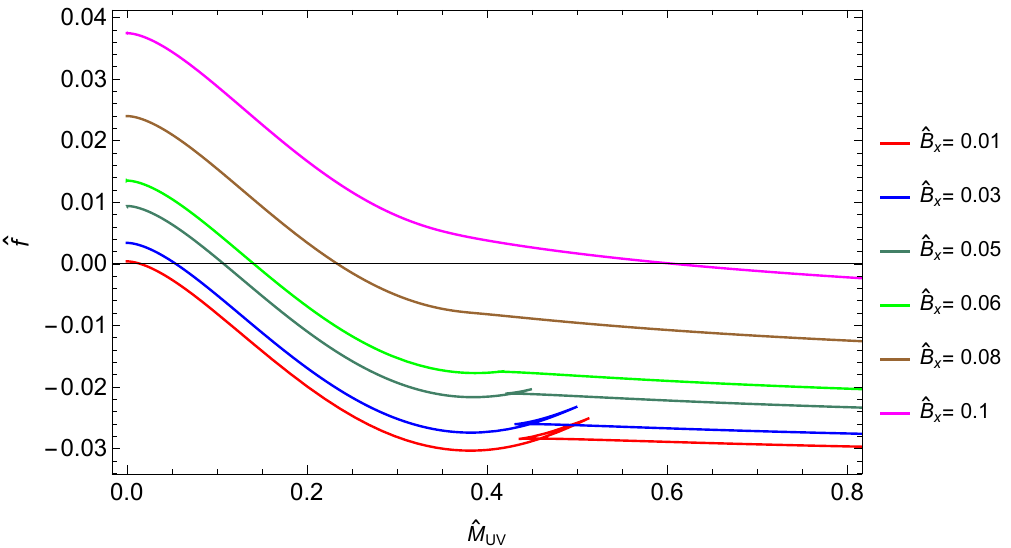}
                                         
        \textit{Figure 3: Free energy of the solutions in Figure 2 against the UV mass showing the disappearance of the swallow tail and first order transition as $\hat{B}_x$ grows.}   
        \end{center}

\begin{center}
\includegraphics[width=6.7cm,height=4.8cm]{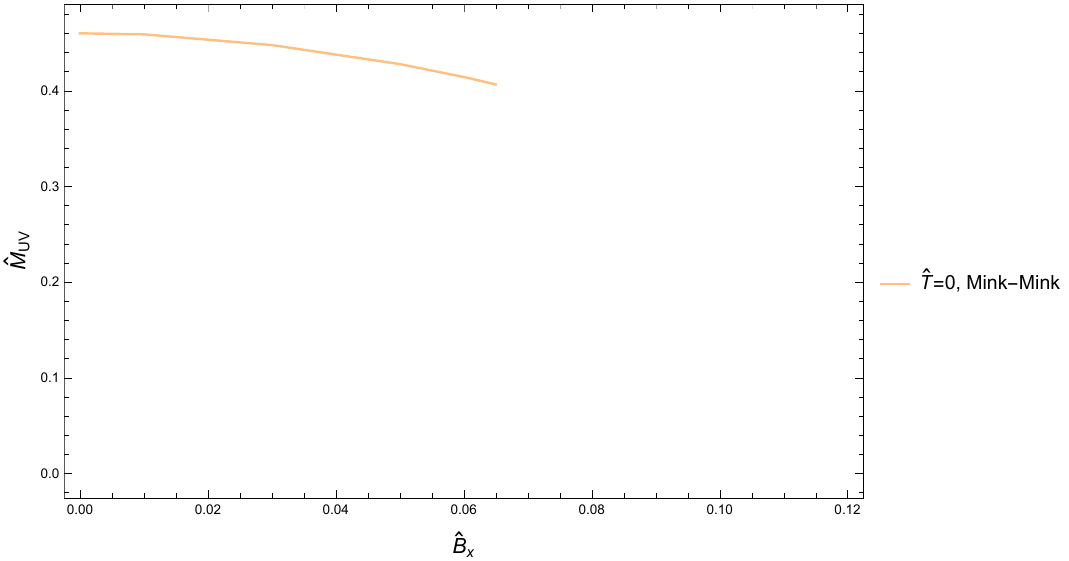}   \includegraphics[width=6.7cm,height=4.8cm]{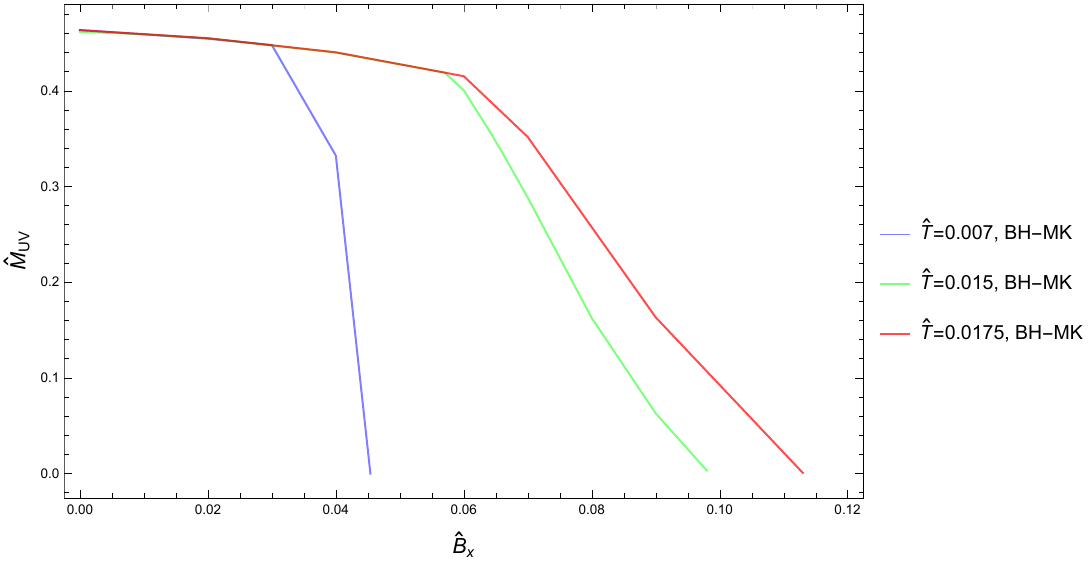}  \\                   
        \textit{Figure 4: The first order phase line in the theory  in the $\hat{M}_{UV}-\hat{B}_{x}$ plane at zero temperature on the left. On the right the thermal phase transition (BH-Mink) at different temperatures.}   
        \end{center}

 form where $\hat{L}'(0) = 0$. In other words, in the presence of $\hat{B}_x$, there is always an infra-red mass generated.

For small $\hat{B}_x$, the first-order behavior remains with a region of UV masses $\hat{M}_{UV}$ where there are multiple solutions. This is unsurprising since when $\hat{B}_x \ll 1$ ($B_x \ll b^2$), the physics at the first-order transition is governed entirely by $b$—only at much lower scales is the massless theory influenced and gapped by $\hat{B}_x$.

As $\hat{B}_x$ rises, though, the transition region shrinks and eventually ceases to exist. There becomes a single unique solution for each UV quark mass. We show this further in the lowest plot in Figure 2, where we plot the UV quark mass against the IR quark mass ($\hat{L}(0)$)—as $\hat{B}_x$ grows, it moves from being multi-valued to single-valued.

In Figure 3 we show the free energy of the solutions after the subtraction of the  counter terms (\ref{ct}). We see the traditional swallow tail structure associated with a first order transition. As $\hat{B}_{x}$ gets larger the structure shrinks before eventually ending. This means that we have encountered a critical end point of the first order transition line in the $\hat{B}_x$ vs $\hat{M}_{UV}$ plane. We plot this phase diagram in Figure 4.

\section{Massless Mode at the Critical Point}

The critical point should correspond to the values of parameters where the free energy between the two competing vacuum states become degenerate and the barrier between those vacua falls to zero height. The Higgs/sigma mode of the symmetry breaking should therefore see a flat direction in the potential at this point and there should be a massless state in the spectrum.

To see this in our system, we look at static fluctuations $\delta \hat{L}$ about the solutions we have determined above, $\hat{L}_0(\hat{\rho})$. Thus in (\ref{Baction}), we write $\hat{L} = \hat{L}_0(\hat{\rho}) + \delta \hat{L}(\hat{\rho}, t)$ and find the linearized equation for $\delta \hat{L}$. We then solve for solutions where $\delta \hat{L} = f(\hat{\rho}) e^{i\hat{E}t}$ and seek solutions $f(\hat{\rho})$ that vanish in the UV and have $f'(0) = 0$ in the IR.

In particular, we will scan lines on the phase diagram varying $\hat{M}_{UV}$, at fixed $\hat{B}_x$, passing across the phase boundary. For example, see Figure 5 for the fluctuation's mass along some trajectories at a fixed $\hat{M}_{UV}$ and varying $\hat{B}_x$. When we cross the phase boundary in this figure, we move from one branch of solutions to the second branch so that the energies of the excitations are always those of the true vacuum. Generically, the scalar mass is larger as $\hat{B}_x$ is larger and as $\hat{M}_{UV}$ is larger, as one would expect. On the other hand, for trajectories that are close to crossing the phase boundary near the critical point, there is a dip in the energy of the excitation as we cross the transition. At the critical point, the mass falls to zero as expected.

\begin{center}
        \includegraphics[width=11cm,height=6cm]{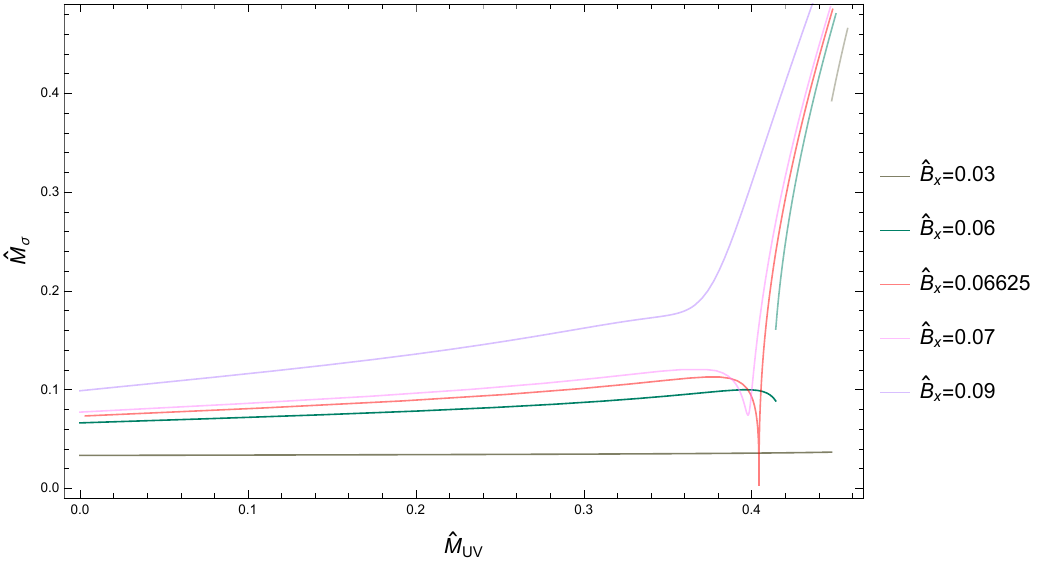}
      
        \textit{Figure 5: The sigma meson mass $\hat{M}_{\sigma}$ against the UV quark mass $\hat{M}_{\rm UV}$ for different $\hat{B}_x$ at $\hat{T}=0$. For values of $\hat{B}_x$ where this trajectory comes near the critical point the sigma mass falls sharply. }   
        \end{center}

Looking for the point where the energy of the excitation is zero is a precise way to identify the position of the critical point. In Figure 6, we show the masses on the two branches of solution on either side of the transition. Here we keep the results in the non-vacuum configurations. Plotting at $\hat{B}_x = 0.065$ and $\hat{B}_x = 0.06625$ shows that the latter is the critical point since the two sets of energies meet at zero.
\vspace{1cm}

        \begin{center}
    \includegraphics[width=8cm,height=5cm]{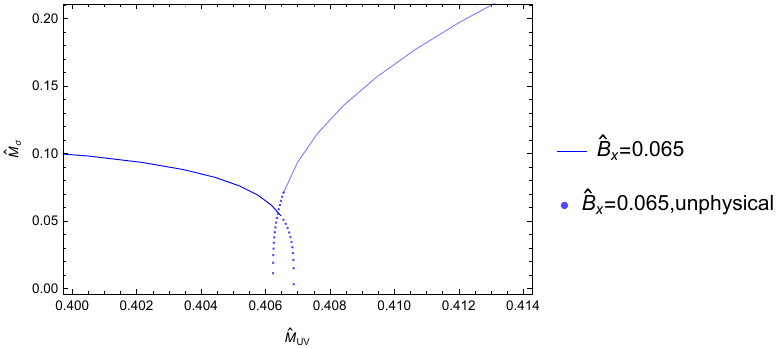}
    \includegraphics[width=7cm,height=5cm]{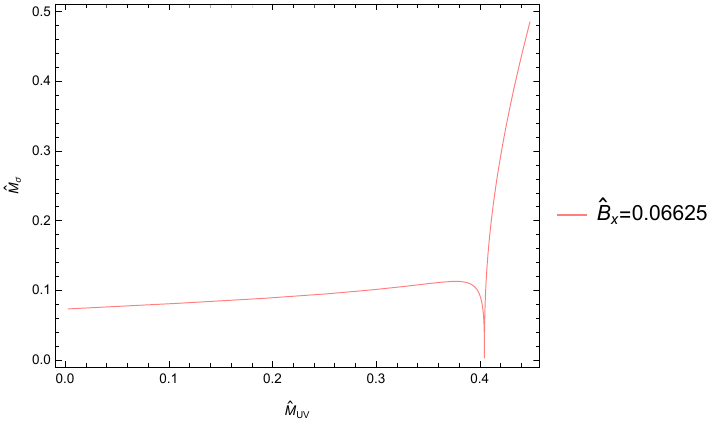} \\
    \textit{Figure 6: $\hat{M}_{\sigma}$ against $\hat{M}_{UV}$ for two different $\hat{B}_x$, $\hat{T}=0$}   
\end{center}

\newpage

\begin{center}
    \includegraphics[width=7cm,height=5cm]{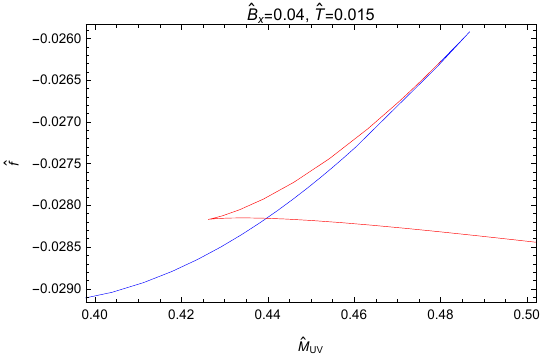}
    \includegraphics[width=7cm,height=5cm]{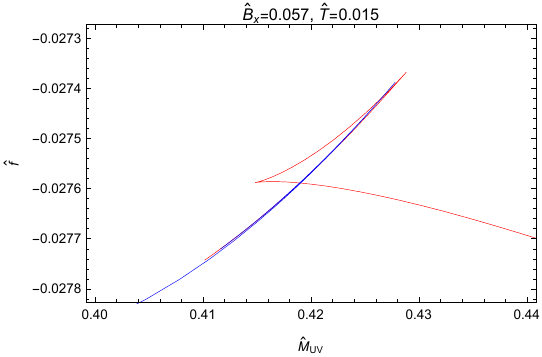}\\
    \includegraphics[width=7cm,height=5cm]{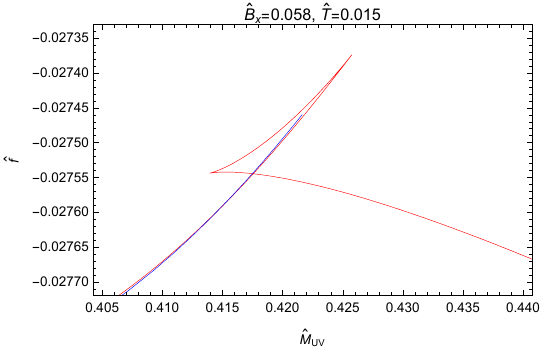}
    \includegraphics[width=7cm,height=5cm]{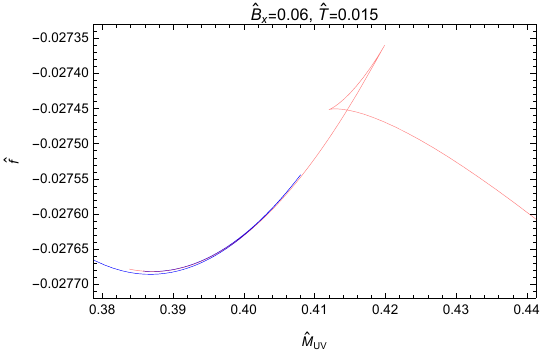}\\
    \textit{Figure 7: Free energy against $\hat{M}_{UV}$ for different $\hat{B}_x$ values at, $\hat{T}=0.0015$. Red line is for Minkowski embedding and the blue line is for black hole embedding.}  
\end{center}

\section{Finite Temperature}

We can consider the effect of temperature on the phase diagram of the Weyl semi-metal with $\hat{B}_x$ in Figure 4 (the zero-temperature phase diagram is on the left, while those on the right show the effects of temperature as we now discuss). To include temperature, we allow $r_H \neq 0$ in (\ref{metricT2}). Moving from zero temperature to infinitesimal temperature introduces a horizon at the origin of the $\hat{L}-\hat{\rho}$ plane. The Weyl semi-metal phase, where the D7 embeddings end at $\hat{\rho} = \hat{L} = 0$, will immediately become a solution that ends on the black hole horizon (this phase has melted quasi-normal mode states rather than stable mesons \cite{Hoyos-Badajoz:2006dzi}). In Figure 4, the black hole phase will extend up the $\hat{M}_{UV}$ axis from the origin up to the phase boundary between the two phases already seen. At any finite $\hat{B}_x > \hat{T}^2$, the magnetic field induces an IR gap, and an infinitesimal horizon will not change the solution—thus, the black hole phase is indeed a line up the $\hat{B}_x = 0$ axis at infinitesimal temperature.

As we increase $\hat{T}$, we can compute the free energy of solutions on vertical lines through Figure 4—at different fixed $\hat{B}_x$, we scan in $\hat{M}_{UV}$. Examples of these plots are shown in Figure 7. At low $\hat{B}_x$, there is a single transition on the transition line we saw before at $\hat{T} = 0$. At higher $\hat{B}_x$ values, two transitions occur—there are two swallowtail regions of the plot, although one is somewhat flattened. One is again the zero-temperature insulator-to-semi-metal transition, whilst the other is a thermal transition where the IR gap on a semi-metal state becomes too small to resist falling onto the horizon of the black hole. The resulting shifting transition boundaries are shown in Figure 4 on the right—the line up the $\hat{M}_{UV}$ axis of the black hole phase at infinitesimal $\hat{T}$ moves out into the $\hat{B}_x-\hat{M}_{UV}$ plane, ending at the critical line of the $\hat{T} = 0$ theory. Eventually, the end of the transition region reaches the critical point on the $\hat{T} = 0$ transition line, and the full transition becomes the usual finite-temperature meson melting transition in the magnetic field theory without $b$ \cite{Albash:2007bk}.

The phase structure at different $T$ value sis shown on the right hand side in Figure 4.
\vspace{1cm}

\section{Bubble Walls}

We now turn to the motion of bubble walls in the theory we have constructed above. The theory we have described allows us to control the parameters of the bubble wall in a holographic setting at zero temperature. In particular, consider Figures 3 and 4 above. In each case we can change the parameters $\hat{M}_{UV}$ and $\hat{B}_x$. In the region with a swallowtail in Figure 3, there are two local minima of the potential for the vacuum configuration (the third solution, for example at fixed $\hat{M}_{UV}$, is the local maximum between the two minima). Thus, by moving in the range of $\hat{M}_{UV}$ of the swallowtail at fixed $\hat{B}_x$, we can find theories with dial-able energy (equal to minus the pressure) differences between the two minima. Now, by changing $\hat{B}_x$, we can move along the line of first-order transitions towards the critical point. At the critical point, the two local minima converge, and the barrier height between them is zero—thus, we can dial the barrier height as well. We sketch our definitions of the energy difference and barrier height in Figure 8  - here we are sketching the effective potential for a theory at fixed $b$, $B_x$ and $M_{UV}$ as we imagine varying the IR quark mass. The turning points correspond to the solutions found on a vertical slice through Figure 3 in the swallow tail region.

\begin{center}
        \includegraphics[width=13cm,height=7cm]{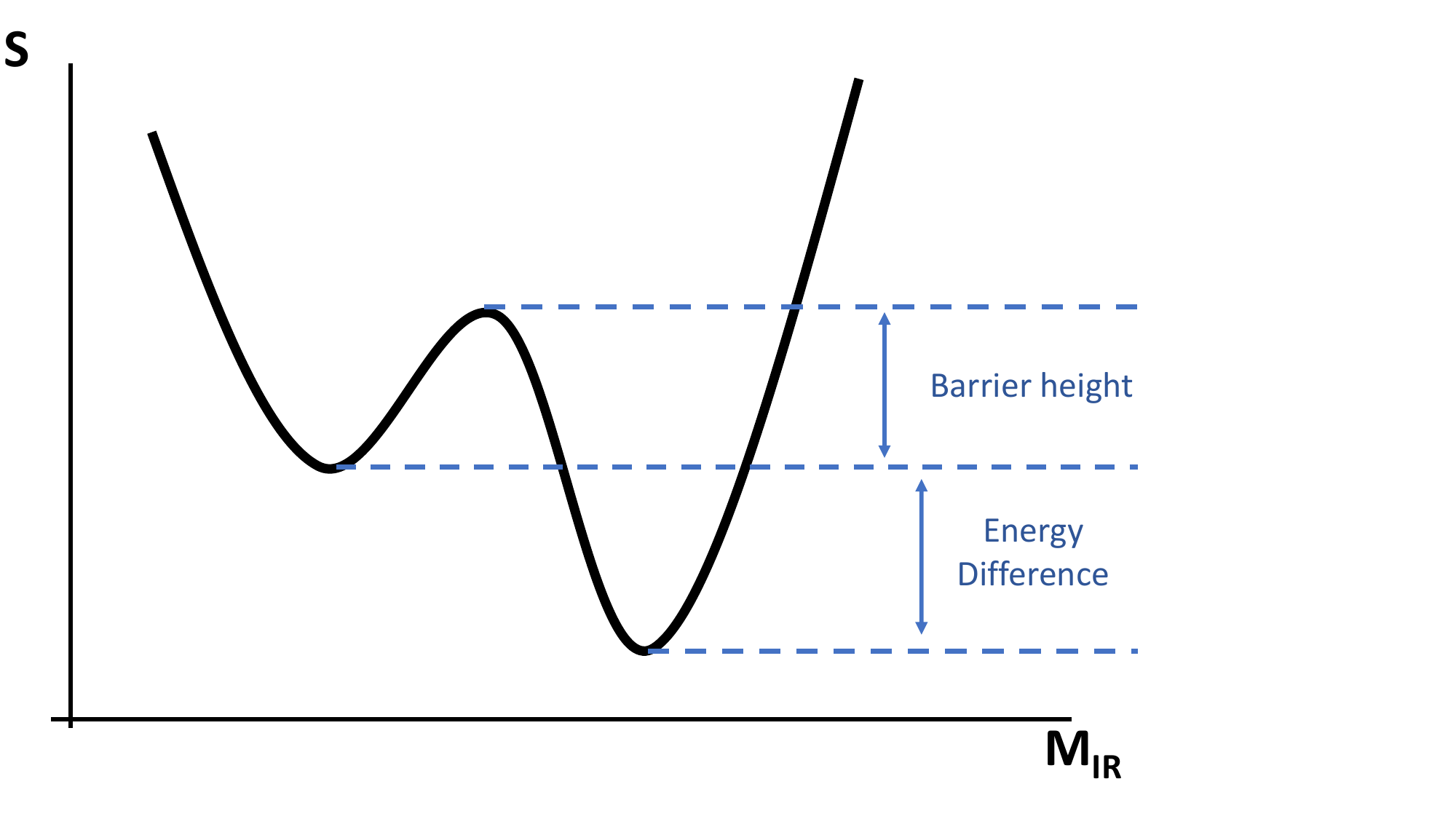}
                                         
        \textit{Figure 8:  A sketch of the effective potential against the IR quark mass showing the parameters we refer to as the barrier height and the energy difference. Here we are in a theory at fixed $b$, $B_x$ and $M_{UV}$. The turning points correspond to the solutions found on a vertical slice through Figure 3 in the swallow tail region.}   
        \end{center}

As a first investigation, we will look at one dimensional bubble walls where the vacuum transitions between the two minima at some value of a spatial direction ($x,y$ or $z$) in these different theories. Remember we have picked $B_x$ and associated the $b$ parameter with the $z$ direction so that in the Weyl semi-metal phase there are massless modes moving in $z$. In fact we have observed no discernible change in behaviours between theories with $x$ and $y$ variation in configurations and will not mention $y$ henceforth.  

We begin by describing a domain wall in the $x$ direction. 
The D7 DBI action with temperature including variation in the
spatial direction $x$ and time dependence is 

\footnotesize
\begin{align}\label{eom:bubble wall in x}
&S_{D7}=-\mathcal{N} \text{vol}\left( \mathbb{R}^{1,3}\right)\int d\rho \mathcal{L}\\
    &\mathcal{L}=\rho^3 g h \sqrt{\left(1+\frac{R^4\,B_x^2}{h^2 (\rho^2+L^2)^2}+\frac{R^4\,b^2 L^2}{h (\rho^2+L^2)^2}\right) \left(1+(\partial_{\rho}L)^2+\frac{R^4 (\partial_{x}L)^2}{h (\rho^2+L^2)^2}-\frac{h}{g^2} \frac{R^4 (\partial_{t}L)^2}{(\rho^2+L^2)^2} \right)}
\end{align}

\normalsize

where $\mathcal{L}$ is the Lagrangian density, and $g=g(r)$ and $h=h(r)$ for short.

The action in terms of rescaled quantities is 

\footnotesize
\begin{align}
&\hat{S}_{D7}=\frac{S_{D7}}{\mathcal{N}R^8}=-\int d\hat{t}\,d\hat{x}\,d\hat{y}\,d\hat{z}\,d\hat{\rho} \, \hat{\mathcal{L}} \\
&\hat{\mathcal{L}}=\hat{\rho}^3 g h \sqrt{\left(1 + \frac{ \hat{B}_x^2}{h^2 (\hat{\rho}^2 + \hat{L}^2)^2} + \frac{ \hat{L}^2}{h (\hat{\rho}^2 + \hat{L}^2)^2}\right) \left(1 + (\partial_{\hat{\rho}} \hat{L})^2 + \frac{ (\partial_{x} \hat{L})^2}{h (\hat{\rho}^2 + \hat{L}^2)^2} -  \frac{h (\partial_{t} \hat{L})^2}{g^2 (\hat{\rho}^2 + \hat{L}^2)^2} \right)}.
\end{align}

\normalsize
For more details, see \autoref{app: rescaled quantities}.

The equation of motion is
\begin{equation}
    -\partial_{\hat{t}} \left( {\partial \hat{\cal L} \over \partial (\partial_{\hat{t} }  \hat{L})}\right) + \partial_{\hat{x}} \left( {\partial \hat{\cal L} \over \partial (\partial_{\hat{x} }  \hat{L})}\right)- {\partial {\hat{L}} \over \partial {\hat{L}}} = 0
\end{equation}

The boundary conditions (6 in total) we will choose are 
\footnotesize
\begin{align}
    &\hat{L}[\hat{\rho},0,\hat{x}] = \hat{L}_0[\hat{\rho},\hat{x}], \label{bc1}\\
    &\partial_{\hat{t}} \hat{L}[\hat{\rho},0,\hat{x}] = 0, \label{bc2}\\
    &\hat{L}[\hat{\rho},\hat{t},-\hat{x}_{\rm max}] = \hat{L}_0[\hat{\rho},-\hat{x}_{\rm max}], \label{bc3}\\
    &\hat{L}[\hat{\rho},\hat{t},\hat{x}_{\rm max}] = \hat{L}_0[\hat{\rho},\hat{x}_{\rm max}],\label{bc4}\\
    &\hat{L}[\hat{\rho}_{uv},\hat{t},\hat{x}] = \hat{L}_0[\hat{\rho}_{uv},\hat{x}], \label{bc5}\\
    &\partial_{\hat{\rho}} \hat{L}[0,\hat{t},\hat{x}] = 0,\label{bc6}
\end{align}
\normalsize
The first boundary condition \eqref{bc1} is simply an ansatz for the initial condition.



Our choice of initial condition for $\hat{L}_0(\hat{\rho},\hat{x})$ is of the form
\footnotesize
\begin{align}\label{eq:initial condition}
    \hat{L}_0(\hat{\rho},\hat{x}) = \frac{1}{2}\left(1 - \frac{2}{\pi}{\rm Arctan}(\hat{x})\right)\hat{L}_1(\hat{\rho}) + \frac{1}{2}\left(1 + \frac{2}{\pi}{\rm Arctan}(\hat{x})\right)\hat{L}_2(\hat{\rho}),
\end{align}
\normalsize
where $\hat{L}_1$ and $\hat{L}_2$ are the $\hat{x}$-independent solutions for the true vacuum and metastable vacuum {\it i.e.} they are explicit solutions of the equations of motion as described in the previous sections of this paper. The Arctan functions simply transition us from one solution to the other as we move in $\hat{x}$ around $\hat{x} = 0$. We plot an example initial condition in Figure 9.

The second boundary condition we chose in \eqref{bc2} dictates that we begin with the initial configuration static.

The third and fourth conditions \eqref{bc3}, \eqref{bc4} assume that the far left and right configuration at the extreme points of the simulation box $\pm \hat{x}_{\rm max}$, away from the bubble 

\begin{center}

\includegraphics[width=6.7cm,height=5.2cm]{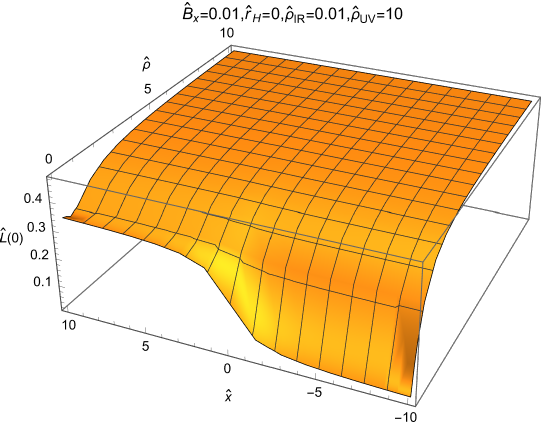}
    
    \textit{Figure 9: $\hat{B}_x=0.01, \hat{M}_{UV}=0.479$. An example of initial conditions for a case where we lie slightly above the phase boundary in Figure 4 where the Weyl semi-metal phase is the metastable vacuum (negative $\hat{x}$) and the true vacuum is the massive state (positive $\hat{x}$).}
\end{center}

  wall ansatz around $\hat{x} = 0$, remain in the vacuum and metastable vacuum configurations throughout the simulation. This of course means one must be careful not to run the configuration for too long in time so that the bubble wall doesn't significantly approach these boundaries.


The boundary condition \eqref{bc5} is essentially ensuring that at the UV boundary all solutions asymptote to the UV choice of the mass term $\hat{M}_{UV}$. This is the case but strictly solutions only reach that value as $\hat{\rho} \rightarrow \infty$ — the condition as written allows one to bring the numerical UV cut off in $\hat{\rho}$ to a finite large value and remain consistent with the initial condition.

Finally, \eqref{bc6} ensures the usual IR regularity condition for Minkowski embeddings that the derivative of $\hat{L}$ vanishes at $\hat{\rho} = 0$ for all $\hat{x}$ and $\hat{t}$.

We now run the time evolution of this configuration using the \verb|NDSolve| function from \verb|MATHEMATICA|. We monitor the energy of the configuration as a function of time to check it does not vary by more than a percent or two from the initial value to monitor the consistency of the solution.  In the simulations we report we use the ranges $\hat{\rho}_{UV}$ between 5 and 10; we vary the spatial direction ranges between 5 and 20; and we vary the time range from 5 to 15 units. These ranges are dictated by \verb|MATHEMATICA|'s ability to produce output and one would like to run to bigger ranges in all parameters. For the examples we give we see basic stability of the solutions' behaviour as we vary ranges in these limits. We stress again, as in the introduction, though that we consider them

\begin{center}
\includegraphics[width=6.7cm,height=4.8cm]{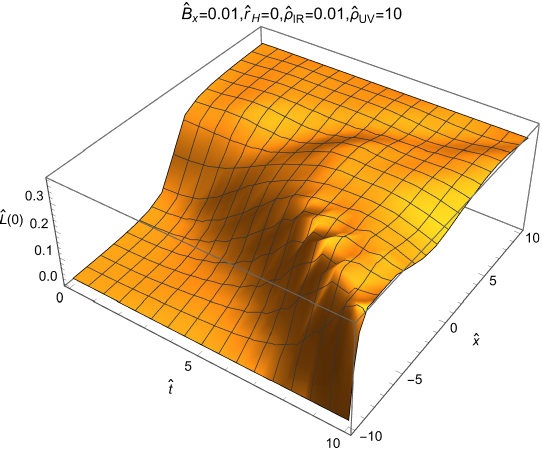} \hspace{0.75cm}
\includegraphics[width=6.7cm,height=4.8cm]{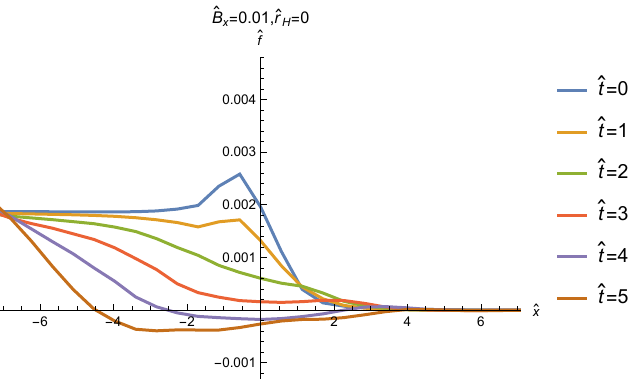}\\
\textit{Figure 10: Time evolution of the initial condition from Fig 9 at $\hat{T}=0$. We show a 3D plot of $\hat{L}(0)$ as a function of $\hat{x}$ and $\hat{t}$. In the second plot, we plot the free energy (minus pressure) against $\hat{x}$ at various times $\hat{t}$. $\hat{M}_{UV} = 0.479$, $\hat{x}_{max}=10,\hat{\rho}_{UV}=10$}
\end{center}

preliminary results 
showing the interesting physics in the set up presented but we would like to see more bespoke computing methods before a systematic survey of the results were made.

For example, we plot the evolution of an initial condition $\hat{L}(0)$ as a function of time and $\hat{x}$ at zero temperature in Fig 10. This is for a case where we have chosen parameters so we lie slightly above the phase boundary in Figure 4, where the Weyl semi-metal phase is the metastable vacuum (negative $\hat{x}$) and the true vacuum is the massive state (positive $\hat{x}$). It clearly shows the bubble wall motion. Note that we cannot also plot 
the $\hat{\rho}$ dependence of the configurations due to the paucity of dimensions. This can be misleading since waves can also propagate along the configuration in $\hat{\rho}$. A solution to this is to plot the action integrated along $\hat{\rho}$ to give the local energy density (or negative pressure) $\hat{f}$ as derived in \autoref{app: rescaled quantities}.  Note here we always add a constant to the action so that the lower of the two vacua is shown at free energy zero - this allows one to easily see the pressure difference between the two vacua when one compares one plot to another below. It is the pressure difference that drives the wall motion. Note that the actual pressure is never therefore unphysically negative in any of the plots we show.  





This we can plot against $\hat{x}$ at different times $\hat{t}$ as in the second plot in Fig 10. We can extract the speed of the bubble wall simply from this graph from the motion of the mid-point.

  The wall's speed, as one might expect in the absence of friction,  has quickly approached the speed of light $c=1$. Finally we note that momentum conservation will ensure there 
  
   \begin{center}   
\includegraphics[width=6.7cm,height=4.8cm]{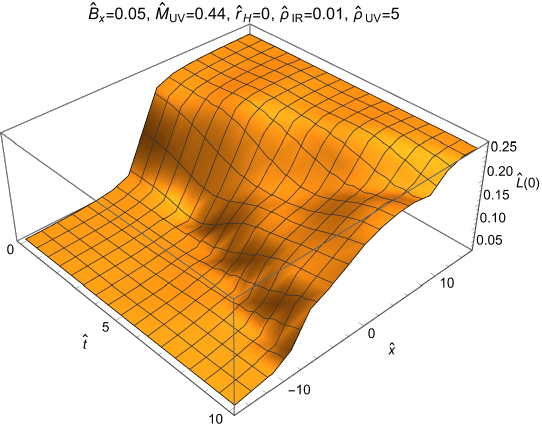} \hspace{0.75cm}
\includegraphics[width=6.7cm,height=4.8cm]{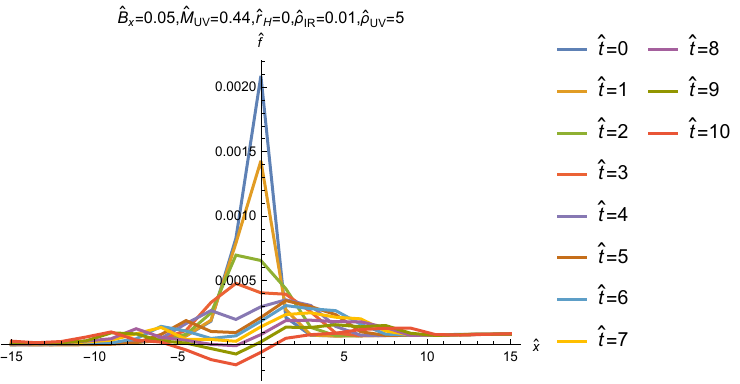}\\
\textit{Figure 11: The time evolution of a configuration with small energy diffrence between the vacuum and metastable vacuum - the energy peak from our ansatz spreads out but it is hard to discern the wall motion in these "slumps". ($\hat{x}_{max}=15, \hat{\rho}_{UV}=5$})\\

\end{center}

  is always a wave propagating in the opposite direction to the motion of the domain wall but here it is a small feature in the energy density plot. Although this is one example of initial conditions the main behaviour is robustly maintained in simulations of other initial conditions we have tried.

We now describe examples of initial conditions where there is additional or different behaviour at zero temperature. 
Our first example is moving our parameter point very close to the phase line so that the difference in energy density between the two sides of the wall is small. For example, choosing $\hat{M}_{UV} = 0.4$ and $\hat{B}_x = 0.05$ gives an energy density difference that is ten times smaller than the previously described case. We show our results for the time evolution in this case in Figure 11. One can see that at $\hat{t} = 0$, the energy density plot is dominated by a spike due to our ansatz for the wall's initial shape. As time progresses, that central energy peak slumps with waves propagating to left and right as the solution collapses to a much wider domain wall, which is apparently preferred in this case. There is little clear motion of the centre of the wall itself though.

Another example of an interestingly different case is to set up the domain wall spatial variation in the $z$ direction rather than $x$, i.e $L(t,\rho,z)$. Here the action is
\footnotesize
\begin{align}
&S_{D7}=-\mathcal{N} \text{vol}\left( \mathbb{R}^{1,3}\right)\int d\rho \mathcal{L}\\
       &\mathcal{L}=\rho^3 g h \sqrt{\left( 1+   \frac{R^4\,B_x^{2}}{h ^2 (\rho^2+L^2)^2}+\frac{R^4\,b^2 L^2}{h(\rho^2+L^2)^2}\right) \left(1+(\partial_\rho L)^2-\frac{R^4\,h (\partial_t L)^2}{g^2(\rho^2+L^2)^2}\right)+\frac{R^4\,(\partial_z L)^2}{h (\rho^2+L^2)^2}}
\end{align} 

\normalsize
where $h=h(r)$ and $g=g(r)$ for short. The action in terms of rescaled quantities is


\normalsize
\begin{center}
\includegraphics[width=6.7cm,height=4.8cm]{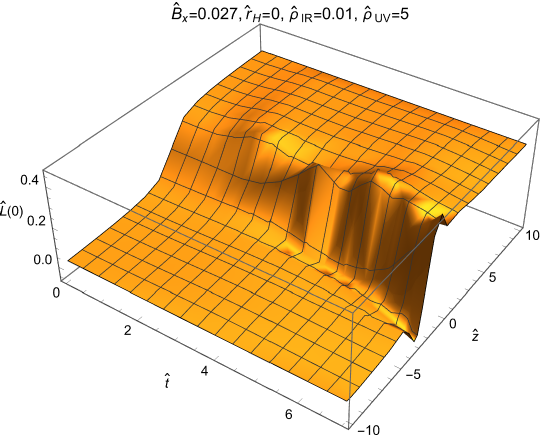}
\hspace{0.75cm} \includegraphics[width=6.7cm,height=4.8cm]{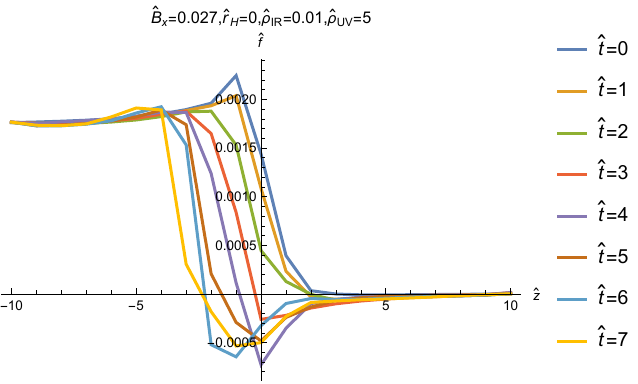}\\
        \textit{Figure 12: An example of a bubble wall moving in the $\hat{z}$ direction which is slowed by the massless modes in the Weyl semi-metal phase. $\hat{B}_x$=0.027, $\hat{r}_H$=0, $\hat{M}_{UV}$=0.476, $\hat{z}_{max}=10, \hat{\rho}_{UV}=5$ - Average speed (from $\hat{t}=0$ to $\hat{t}=7$)=0.52c. }  
\end{center}

\footnotesize
\begin{align}
&S_{D7} =\frac{S_{D7}}{\mathcal{N}R^8} =-\int d\hat{t} d\hat{x} d\hat{y} d\hat{z} d\hat{\rho} \,\hat{\mathcal{L}} \\
&\mathcal{L} = \hat{\rho}^3 g h \sqrt{\left( 1 + \frac{\hat{B}_x^{2}}{h^2 (\hat{\rho}^2 + \hat{L}^2)^2} + \frac{\hat{L}^2}{h (\hat{\rho}^2 + \hat{L}^2)^2} \right) \left(1 + (\partial_{\hat{\rho}} \hat{L})^2 - \frac{h (\partial_{\hat{t}} \hat{L})^2}{g^2 (\hat{\rho}^2 + \hat{L}^2)^2} \right) + \frac{(\partial_{\hat{z}} \hat{L})^2}{h (\hat{\rho}^2 + \hat{L}^2)^2}}.
\end{align}
\normalsize

whereas the initial condition and the boundary condition are set up in $\hat{z}$ direction:
\begin{align}
    &\hat{L}[\hat{\rho}, 0, \hat{z}] = \hat{L}_0[\hat{\rho}, \hat{z}], \\
    &\partial_{\hat{t}} \hat{L}[\hat{\rho}, 0, \hat{z}] = 0, \\
    &\hat{L}[\hat{\rho}, \hat{t}, -\hat{z}_{\rm max}] = \hat{L}_0[\hat{\rho}, -\hat{z}_{\rm max}], \\
    &\hat{L}[\hat{\rho}, \hat{t}, \hat{z}_{\rm max}] = \hat{L}_0[\hat{\rho}, \hat{z}_{\rm max}], \\
    &\hat{L}[\hat{\rho}_{uv}, \hat{t}, \hat{z}] = \hat{L}_0[\hat{\rho}_{uv}, \hat{z}], \\
    &\partial_{\hat{\rho}} \hat{L}[0, \hat{t}, \hat{z}] = 0.
\end{align}

For these set ups we systematically find the bubble wall motion to be slower at around $v=0.5\,c$ over the timescales that we run the simulation, see and example in Figure 12. Here the bubble wall is moving into a region with light quarks moving in $\hat{z}$. These results suggests that those light quarks moving in $\hat{z}$ provide friction to the motion in $\hat{z}$. Since these are not a traditional thermal bath causing friction (relevant to cosmology) we will not further investigate this here although it is an interesting observation.

\begin{center}
\includegraphics[width=6.8cm,height=4.2cm]{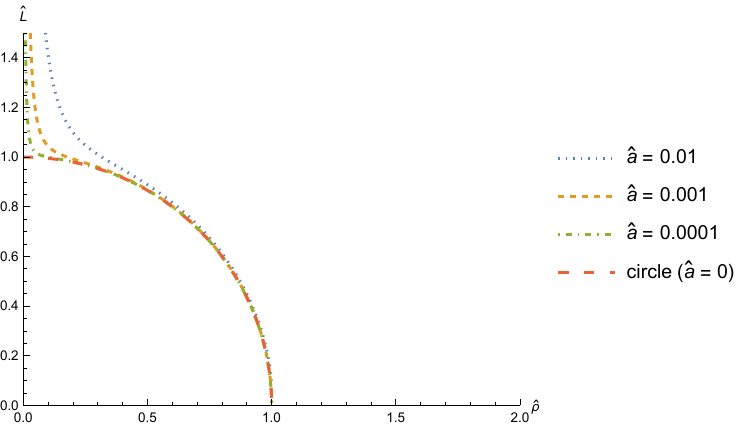}\\
    \textit{Figure 13: The boundary in (\ref{eq:profile}) where we set initial conditions in the $\hat{L}-\hat{\rho}$ plane - here $\hat{c}=1$, $\hat{a}=0,0.01,0.001,0.0001$.}
\end{center} 
\section{Bubble Walls at Finite Temperature}



To study the effect of a thermal bath on the bubble wall motion, we will include non-zero \( r_H \) in \eqref{metricT2}. See Appendix A to move to the rescaled coordinates leading to  $\hat{r}_H$ in units of $b$. It is helpful to turn the space in the \( \hat{L}-\hat{\rho} \) plane outside the black hole into a square coordinate grid. 
To do this, we define the IR edge of that space as living on the curve
\footnotesize
\begin{align}
  f(\hat{\rho}) \equiv \hat{L}(\hat{\rho}) = \sqrt{\hat{c}^2 - \hat{\rho}^2 \left(1 - \frac{\hat{a}}{\hat{\rho}^4}\right)},
  \label{eq:profile}
\end{align}
\normalsize
where \( \hat{c} \) is the BH radius, and \( \hat{a} \) plays the role of a cutoff away from the horizon. 
If \( \hat{a} \) is infinitesimal, it approaches the Minkowski and black hole boundaries, as shown in Figure 13.

To make the space ``square," we move to the new coordinate system
\footnotesize
\begin{align}
    &\tilde{L} = \hat{L}, \\
    &\tilde{\rho} = \hat{\rho} - \frac{ \sqrt{ \hat{c}^2 - \hat{L}^2 \pm \sqrt{4 \hat{a} + \left(\hat{c}^2 - \hat{L}^2\right)^2}}}{\sqrt{2}},
    \label{eq:rho_tilde}
\end{align}
\normalsize

where we choose the plus sign to make $\Tilde{\rho}$ real and positive.

We now recompute the DBI action \eqref{Equation: DBI action and WZ action} in the new coordinates.  We obtain the induced metric of the D7 branes (according to \ref{rescaled metric}):
\begin{align}
    \begin{split}
        ds^2 = & \, \hat{g}_{\hat{t}\hat{t}} \, d\hat{t}^2 + \hat{g}_{\hat{x}\hat{x}} \, d\hat{x}^2 + \hat{g}_{\hat{y}\hat{y}} \, d\hat{y}^2 + \hat{g}_{\hat{z}\hat{z}} \, d\hat{z}^2 \\
               & + \hat{g}_{\tilde{\rho}\tilde{\rho}} \, d\tilde{\rho}^2 + \hat{g}_{33} \, ds^2_{S^3}
    \end{split}
\end{align}
where
\footnotesize
\begin{equation}
    \hat{g}_{\hat{t}\hat{t}} = -R^2\frac{\left((f(\tilde{L}) + \tilde{\rho})^2 + \tilde{L}^2\right) \tilde{g}^2}{ \tilde{h}}
\end{equation}

\begin{equation}
    \hat{g}_{\hat{x}\hat{x}} = \hat{g}_{\hat{y}\hat{y}} = R^2\left((f(\tilde{L}) + \tilde{\rho})^2 + \tilde{L}^2\right) \tilde{h} 
\end{equation}

\begin{equation}
    \begin{array}{ccl}
        \hat{g}_{\hat{z}\hat{z}} &=& \hat{g}_{\hat{z}\hat{z}} + \hat{g}_{\hat{\phi}\hat{\phi}} = R^2\left((\hat{\rho}^2 + \hat{L}^2) * h  +   \frac{\hat{L}^2}{\hat{\rho}^2 + \hat{L}^2}\right) \\
        &=&R^2 \left( \left((\tilde{\rho} + f(\tilde{L}))^2 + \tilde{L}^2\right) * \tilde{h}   + \frac{\tilde{L}^2}{(\tilde{\rho} + f(\tilde{L}))^2 + \tilde{L}^2}\right)
    \end{array}
\end{equation}

\begin{equation}
    \begin{array}{ccl}
        \hat{g}_{\tilde{\rho}\tilde{\rho}} &=& \hat{g}_{\hat{\rho}\hat{\rho}} \frac{\partial \hat{\rho}}{\partial \tilde{\rho}} \frac{\partial \hat{\rho}}{\partial \tilde{\rho}} + \hat{g}_{\hat{L}\hat{L}} \frac{\partial \hat{L}}{\partial \tilde{\rho}} \frac{\partial \hat{L}}{\partial \tilde{\rho}} \\
        &=& \hat{g}_{\hat{\rho}\hat{\rho}} \frac{\partial \hat{\rho}}{\partial \tilde{\rho}} \frac{\partial \hat{\rho}}{\partial \tilde{\rho}} + \hat{g}_{\hat{L}\hat{L}} \frac{\partial \tilde{L}}{\partial \tilde{\rho}} \frac{\partial \tilde{L}}{\partial \tilde{\rho}} \\
        &=& \frac{R^2}{\hat{\rho}^2 + \hat{L}^2} \left( \frac{\partial \hat{\rho}}{\partial \tilde{\rho}} \frac{\partial \hat{\rho}}{\partial \tilde{\rho}} + \frac{\partial \tilde{L}}{\partial \tilde{\rho}} \frac{\partial \tilde{L}}{\partial \tilde{\rho}} \right) \\
        &=& R^2\frac{\left(\tilde{L}'(\tilde{\rho}) f'(\tilde{L}) + 1\right)^2 + \tilde{L}'(\tilde{\rho})^2}{(f(\tilde{L}) + \tilde{\rho})^2 + \tilde{L}^2}
    \end{array}
\end{equation}

\begin{equation}
    \hat{g}_{33}
    = \frac{R^2 \, \hat{\rho}^2}{\hat{L}^2 + \hat{\rho}^2} 
    = \frac{R^2 (f(\tilde{L}) + \tilde{\rho})^2}{(f(\tilde{L}) + \tilde{\rho})^2 + \tilde{L}^2}
\end{equation}

\begin{equation}
    \tilde{g} = 1 - \frac{\hat{r}_H^4}{\left((f(\tilde{L}) + \tilde{\rho})^2 + \tilde{L}^2\right)^2}
\end{equation}

\begin{equation}
    \tilde{h} = 1 + \frac{\hat{r}_H^4}{\left((f(\tilde{L}) + \tilde{\rho})^2 + \tilde{L}^2\right)^2}
\end{equation}

\begin{equation}
    \frac{\partial \hat{\rho}}{\partial \tilde{\rho}} = \frac{\partial}{\partial \tilde{\rho}} \left( \tilde{\rho} + f(\tilde{L}(\tilde{\rho})) \right) = 1 + f'(\tilde{L}) \partial_{\tilde{\rho}} \tilde{L}
\end{equation}
\normalsize


In the presence of background magnetic field $\hat{B}_x$, we find the DBI action 
\footnotesize
\begin{equation}
    \begin{array}{ccl}\label{eq:actionTilde}
    S_{D7}&=&-\mathcal{N}\text{vol}(\mathbb{R}^{1,3}) R^8 \int d\tilde{\rho} \hat{\mathcal{L}}\\
    \hat{\mathcal{L}}& =& \left(f(\tilde{L})+\tilde{\rho }\right)^3 \Tilde{g} \Tilde{h} 
        \sqrt{\left(\tilde{L}' f'(\tilde{L})+1\right)^2+\tilde{L}'^2} \nonumber\\ & & \\ & & \left. \right. \hspace{3cm}\times \sqrt{1+\frac{\tilde{L}^2}{\Tilde{h}\left(\left(f(\tilde{L})+\tilde{\rho }\right)^2+\tilde{L}^2\right)^2}+\frac{ B_x^2}{\Tilde{h}^2\left(\left(f(\tilde{L})+\tilde{\rho }\right)^2+\tilde{L}^2\right)^2}},
    \end{array}
\end{equation}

\begin{center}
\begin{align}
&\includegraphics[width=7.5cm,height=5cm]{rescaled_fig10b.pdf} \hspace{0.75cm}
\includegraphics[width=6.7cm,height=5cm]{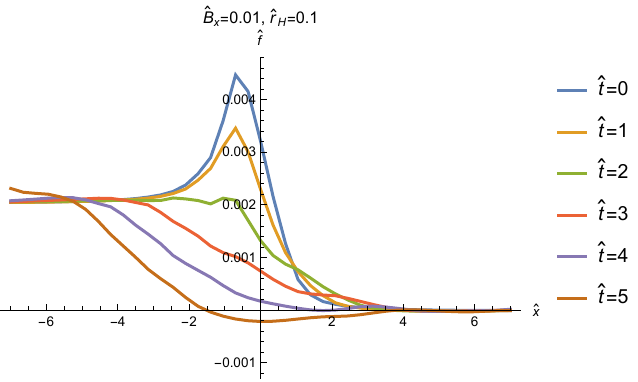}
\end{align}
\textit{Fig 14: Two configurations of bubble wall at $\hat{B}_x=0.01$ with the same pressure difference at $\hat{r}_H=0$ and $\hat{r}_H=0.1$. (Left: $\hat{M}_{UV}=0.479$, $\hat{x}_{max}=10, \hat{\rho}_{UV}=10$; Right: $\hat{M}_{UV}=0.446$, $\hat{x}_{max}=7,  \hat{\rho}_{UV}=5$). The thermal bath impedes the bubble wall motion.}
\end{center}   \vspace{-0.4cm}

\normalsize
where the rescaled free energy (- pressure) density is defined 
\footnotesize
\begin{align}
    \hat{f}=  \int \tilde{\rho} \hat{\mathcal{L}}.
\end{align}
\normalsize

Now we can solve the resulting equation of motion for $\tilde{L}(\tilde{\rho})$. We need to pick boundary conditions at $\tilde{\rho}=0$. In the Minkowski region 
we should continue to pick $\tilde{L}'(\tilde{\rho})=0$.

In fact we choose this same condition at all $\tilde{\rho}$. The reason is that near the black hole horizon the solutions have an attractor behaviour and are very insensitive to the boundary condition choice \cite{Babington:2003vm} presumably reflecting that in-falling matter takes an infinite time to penetrate the horizon from the view of an external observer. 

We can now present two example simulations that display the behaviours we observe with temperature. Firstly we can study the configuration of Figure 10 which at zero temperature was a clear cut wall that accelerated quickly to move at $c$. In Figure 14 we reproduce the plot we had of pressure against position at varying times for that wall (on the left). We also show the result for as large a temperature as we can achieve with the same pressure difference across the wall (we adjust $\hat{M}_{UV}$ to keep this constant as $\hat{T}$ rises). The main result is that the speed of the wall, at least in the initial phase of acceleration, is slowed to $0.7c$ showing the presence of friction from the thermal bath. Note the pressure plots are not identical at $t=0$ with the thermal configuration having more energy stored in the wall - it is possible to adjust this by adding a further parameter writing ${\rm ArcTan}(\gamma \hat{x})$ in our initial ansatz (\ref{eq:initial condition}) and varying

\begin{center}
\begin{align}
&\includegraphics[width=6.7cm,height=5cm]{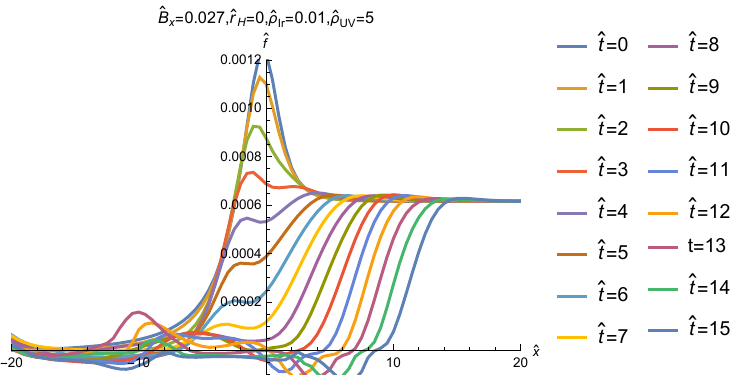}
&\includegraphics[width=6.7cm,height=5cm]{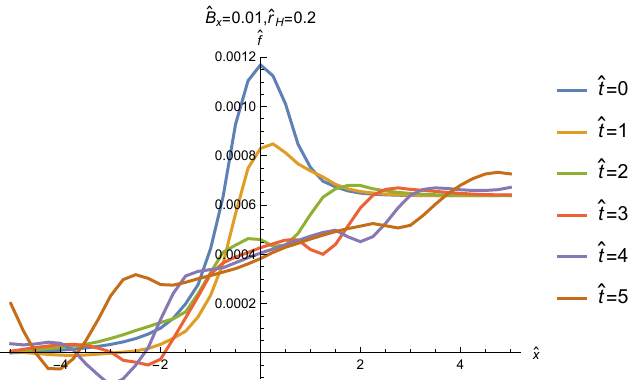}
\end{align}
\textit{Fig 15: A bubble wall configuration - the two figures show walls with the same pressure difference  at  $\hat{r}_H=0$ and $\hat{r}_H=0.2$. (Left: $\hat{M}_{UV}=0.415$, $\hat{x}_{max}=20$, $\hat{\rho}_{UV}=5$; Right: $\hat{M}_{UV}=0.456$,$\hat{x}_{max}=5$, $\hat{\rho}_{UV}=7$) The finite temperature case shows a double step evolution. }
\end{center}

$\gamma$. The conclusion does not change as to the behaviour though when we run those simulations.

The second example we exhibit is shown in Figure 15. Here we look at a configuration with lower pressure difference. At $\hat{T}=0$ the evolution goes through an initial slump before developing into a moving domain wall. This is shown on the left of Figure 15. On the right we show a configuration with the same pressure difference but evolving at finite $\hat{T}=0.2$. We have not been able to evolve the configuration for as long at finite

$\hat{T}$ as at $\hat{T}=0$ but the early stages show a possible new behaviour distinct from simply an accelerating wall. Here the configuration shows two pressure waves moving in opposite directions with a pressure plateau between them. This is possibly a configuration that is intermediate to the behaviour described in \cite{Janik:2022wsx} where the pressure was seen to fully equilibriate around the wall with a separate back pressure wave moving in the opposite direction. The figure shown is the most controlled example we have seen of this behaviour. It would be interesting to fully explore the parameter space in the future to see how these walls evolve at different parameter choices. For the moment this is beyond the Mathematica analysis we present here.

\section{Conclusion}

The goal of this paper was to find a holographic description of a strongly coupled system where we could control bubble wall parameters. In particular in the D3/probe D7 systems we have constructed a line of first order transitions ending at a critical point. To do this we applied a backgound axial vector field vev $b$ which induces a massless Weyl semi-metal phase and a magnetic field $B_x$ that generates mass to a massive quark system. We have studied the phase structure of this theory revealing the zero temperature and finite temperature phase diagrams. 

At zero temperature we have solved PDEs for the motion of a one dimensional bubble wall showing it accelerate to the speed of light. We have also developed methods to allow us to add in a thermal bath. The resulting simulations showed new behaviours beyond the zero temperature results including slowing friction and a double wall structure in cases with smaller pressure differences.

The strongest result after these simulations is simply that the programme outlined here indeed gives us the ability to study domain wall motion in varying thermal baths at strong coupling. Our results with a thermal bath here have been preliminary because we have allowed ourselves to depend on the \verb|MATHEMATICA| \verb  |NDSolve|  black box  which in the future we hope to improve upon. However, we hope this study will inspire the study of bubble walls in other holographic settings to try to understand how generic the behaviours we see are in strongly coupled systems. Were gravitational wave signatures of a transition to be observed in experiment it may become interesting to have these descriptions of the strongly coupled cases. 

\bigskip \noindent {\bf Acknowledgements:}  
The authors are grateful for many early discussions with Andy O'Bannon. We thanks Ronnie Rodgers for comments also. NE's work was supported by the
STFC consolidated grants ST/T000775/1 and ST/X000583/1.

\newpage

\appendix
\section{Appendix A: Rescaled Quantities}\label{app: rescaled quantities}

Here we summarize the rescaled quantities we use in the numerical analysis.  We write the dimensions of $L$ and $\rho$ in units of $R^2b$ 
\begin{equation}\label{eqns:scale_1}
    \rho = R^2\,b\, \hat{\rho},~~~~~L = R^2\,b\, \hat{L},~~~~~
    r = R^2\,b\, \hat{r},~~~~~
    r_H = R^2\,b\, \hat{r}_H,~~~~~
    T = R^2\,b\, \hat{T}
\end{equation}
where $\hat{r}^2=\hat{\rho}^2+\hat{L}^2$.
According to the asymptotic expansion of $L$ \eqref{asymptotic expansion of L}, we can defined the rescaled UV mass
$M_{UV}=R^2\,b\, \hat{M}_{UV}.$

Then, $g(r)$, $h(r)$, $L'(\rho)$ are rescaled as
\begin{align}
    g(r)=1-\frac{r_H^4}{r^4}=1-\frac{\hat{r}_H^4}{\hat{r}^4}=g(\hat{r}),~~~~~~
    h(r)=1+\frac{r_H^4}{r^4}=1+\frac{\hat{r}_H^4}{\hat{r}^4}=h(\hat{r}),\\
    L'(\rho)=\frac{\partial L}{\partial \rho}=\frac{\partial \hat{L}}{\partial \hat{\rho}} =\hat{L}'(\hat{\rho}).
\end{align}

The magnetic fields $B_x$, $B_y$, $B_z$ are rescaled as
\begin{align}
    B_x=R^2 b^2 \hat{B}_x,~~~~~
    B_y=R^2 b^2 \hat{B}_y,~~~~~
    B_z=R^2 b^2 \hat{B}_z.
\end{align}
Substituting the rescaled field back into the DBI action \eqref{eqns:DBI action without B}, we obtained the DBI action in terms of rescaled quantities
\begin{align}
    S_{D7}=-\mathcal{N} \text{vol}\left( \mathbb{R}^{1,3}\right)(R^2b)^4\int d\hat{\rho}\, \hat{\rho}^3  \sqrt{\left( 1+\frac{\hat{L}^2}{(\hat{\rho}^2+\hat{L}^2)^2}\right)(1+\hat{L}'^2)}
\end{align}
where $\hat{L}'=\hat{L}'(\hat{\rho})$, $\text{vol}\left( \mathbb{R}^{1,3}\right)=\int dt dx dy dz$

In the section on bubble walls, the field $L$ also depends on the spacetime coordinates, i.e. $t,x,y,z$, and the space time coordinates are written in units of $1/b$, i.e.
\begin{align}
    t = \frac{1}{b}\hat{t}, \quad
    x = \frac{1}{b}\hat{x}, \quad
    y = \frac{1}{b}\hat{y}, \quad
    z = \frac{1}{b}\hat{z}
\end{align}
The metric \eqref{metricT2} becomes
\footnotesize
\begin{align}\label{rescaled metric}
    ds^2 &= R^4  \frac{\hat{r}^2}{R^2} \left( -\frac{g^2(\hat{r})}{h(\hat{r})} d\hat{t}^2 + h(\hat{r}) (d\hat{x}^2 + d\hat{y}^2 + d\hat{z}^2) \right) + \frac{R^2}{\hat{r}^2} \left( d\hat{\rho}^2 + \hat{\rho}^2 ds^{2}_{S^3} + d\hat{L}^2 + \hat{L}^2 d\phi^{2} \right)\\
    &= R^2  \left(\hat{r}^2 \left( -\frac{g^2(\hat{r})}{h(\hat{r})} d\hat{t}^2 + h(\hat{r}) (d\hat{x}^2 + d\hat{y}^2 + d\hat{z}^2) \right) + \frac{1}{\hat{r}^2} \left( d\hat{\rho}^2 + \hat{\rho}^2 ds^{2}_{S^3} + d\hat{L}^2 + \hat{L}^2 d\phi^{2} \right)\right)\end{align}
\normalsize

The original action from the DBI action is
\footnotesize
\begin{align}
&S_{D7}=-\mathcal{N} \int dt\,dx\,dy\,dz\, d\rho \mathcal{L}\\
    &\mathcal{L}=\rho^3 g h \sqrt{\left(1+\frac{R^4\,B_x^2}{h^2 (\rho^2+L^2)^2}+\frac{R^4\,b^2 L^2}{h (\rho^2+L^2)^2}\right) \left(1+(\partial_{\rho}L)^2+\frac{R^4 (\partial_{x}L)^2}{h (\rho^2+L^2)^2}-\frac{h}{g^2} \frac{R^4 (\partial_{t}L)^2}{(\rho^2+L^2)^2} \right)},
\end{align}
\normalsize
and after rescaling, it becomes
\footnotesize
\begin{align}
&S_{D7} = -\mathcal{N} \, (R^2 b)(1/b)^4 \int d\hat{t}\,d\hat{x}\,d\hat{y}\,d\hat{z}\,d\hat{\rho} \, \mathcal{L} \\
&\mathcal{L} = (R^2b)^3\hat{\rho}^3 g h \sqrt{\left(1 + \frac{ \hat{B}_x^2}{h^2 (\hat{\rho}^2 + \hat{L}^2)^2} + \frac{ \hat{L}^2}{h (\hat{\rho}^2 + \hat{L}^2)^2}\right) \left(1 + (\partial_{\hat{\rho}} \hat{L})^2 + \frac{ (\partial_{x} \hat{L})^2}{h (\hat{\rho}^2 + \hat{L}^2)^2} - \frac{h}{g^2} \frac{ (\partial_{t} \hat{L})^2}{(\hat{\rho}^2 + \hat{L}^2)^2} \right)},
\end{align}
\normalsize
and hence we can redefine the rescaled action and  Lagrangian density
as
\footnotesize
\begin{align}
&\hat{S}_{D7}=\frac{S_{D7}}{\mathcal{N}R^8}=-\int d\hat{t}\,d\hat{x}\,d\hat{y}\,d\hat{z}\,d\hat{\rho} \, \hat{\mathcal{L}} \\
&\hat{\mathcal{L}}=\hat{\rho}^3 g h \sqrt{\left(1 + \frac{ \hat{B}_x^2}{h^2 (\hat{\rho}^2 + \hat{L}^2)^2} + \frac{ \hat{L}^2}{h (\hat{\rho}^2 + \hat{L}^2)^2}\right) \left(1 + (\partial_{\hat{\rho}} \hat{L})^2 + \frac{ (\partial_{x} \hat{L})^2}{h (\hat{\rho}^2 + \hat{L}^2)^2} - \frac{h}{g^2} \frac{ (\partial_{t} \hat{L})^2}{(\hat{\rho}^2 + \hat{L}^2)^2} \right)}
\end{align}
\normalsize
Then for $L(\rho,t,x)$, the action becomes
\begin{align}  \hat{S}_{D7}=\frac{S_{D7}}{\mathcal{N}R^8}=-\hat{\text{vol}}\left( \mathbb{R}^{2}\right)\int d\hat{t}\,d\hat{x}\,d\hat{\rho} \, \hat{\mathcal{L}}
\end{align}
where $\hat{\text{vol}}\left( \mathbb{R}^{2}\right)=\int d\hat{y}\,d\hat{z}$ and we can define the rescaled free energy density $\hat{f}$ and rescaled pressure $\hat{P}$ as
\begin{align}\label{rescaled free energy density}
    -\hat{P}=\hat{f}=-\frac{S_{D7}}{\mathcal{N} R^4 \text{vol}\left( \mathbb{R}^{2}\right)}=\int d \hat{\rho} \,\hat{\mathcal{L}}
\end{align}
In analogy, the rescaled free energy density and rescaled pressure can be defined in the same form 
\begin{align}
    -\hat{P}=\hat{f}=-\frac{S_{D7}}{\mathcal{N} R^4 \hat{\text{vol}}\left( \mathbb{R}^{2}\right)}=\int d \hat{\rho} \,\hat{\mathcal{L}},
\end{align}
where $\hat{\text{vol}}\left( \mathbb{R}^{2}\right)=\int d\hat{x}\,d\hat{z}$ for $L(\rho,t,y)$, $\hat{\text{vol}}\left( \mathbb{R}^{2}\right)=\int d\hat{x}\,d\hat{y}$ for $L(\rho,t,z)$.

\textbf{The Wess-Zumino term $S_{WZ}$ will be non-zero when $B_z\neq0$:}
\footnotesize
\begin{align}
&\frac{1}{2}N_f T_{D7} \int \frac{2R^4\rho^4}{(L(\rho)^2+\rho^2)^2}\phi'(z)(B_z (\partial_\rho A_t -\partial_t A_\rho)-\partial_y A_\rho\, \partial_x A_t+\partial_y A_t \partial_x A_\rho) dt\wedge 
\\ & \left. \right. \hspace{3cm} dx \wedge dy \wedge dz \wedge d\rho \wedge d\beta_1 \wedge d\beta_2 \wedge d\beta_3 \nonumber \\
&=\frac{1}{2}N_f T_{D7} \int dx^8\,\frac{2R^4\rho^4}{(L(\rho)^2+\rho^2)^2}\phi'(z)B_z\, \partial_\rho A_t \end{align}
\textbf{where we choose radial gauge $A_\rho=0$, and $dx^8=dtdxdydzd\rho d\beta_1 d\beta_2 d\beta_3$}
\normalsize

\bibliographystyle{utphys}
\bibliography{main.bib}
\end{document}